\documentclass[aps,prb,twocolumn,superscriptaddress,no-footinbib,floatfix,amsmath,amssymb]{revtex4-1}

\usepackage{graphicx}
\usepackage{dcolumn}
\usepackage{bm}
\usepackage[normalem]{ulem}
\usepackage{subfigure} 
\usepackage{soul}
\usepackage[table]{xcolor}
\usepackage{color}
\usepackage{siunitx}
\usepackage{amsmath}
\usepackage{amsfonts}
\usepackage{amssymb}
\usepackage{textcomp}
\setstcolor{red}

\usepackage[utf8]{inputenc}
\usepackage[T1]{fontenc}
\usepackage{mathptmx}

\usepackage{bbm}

\DeclareMathAlphabet{\mathpzc}{OT1}{pzc}{m}{it}

	\renewcommand{\exp}[1]{e^{#1}}

	\newcommand{\kk}{\mathbf{k}}
	\newcommand{\qq}{\mathbf{q}}
	\newcommand{\QQ}{\mathbf{Q}}

	\newcommand{\pp}{\mathbf{p}}
	\newcommand{\PP}{\mathbf{P}}
	\newcommand{\GG}{\mathbf{G}}
	\newcommand{\KK}{\mathbf{K}}
	
	\newcommand{\RR}{\mathbf{R}}

	\newcommand{\rr}{\mathbf{r}}
	\newcommand{\xxi}{\boldsymbol{\xi}}
	
	\newcommand{\rrho}{\boldsymbol{\rho}}
	
	\newcommand{\ket}[1]{\big| #1 \big>}

	\newcommand{\braoket}[3]{\left<  #1 \left| #2 \right| #3 \right>}

	\newcommand{\mm}{\mathpzc{M}}
	\newcommand{\site}{\mathpzc{S}}

\begin{document}

\title{Theory of moir\'e localized excitons in transition-metal dichalcogenide heterobilayers}

\author{David A.\ Ruiz-Tijerina}
\email{david.ruiz-tijerina@cnyn.unam.mx}
\affiliation{Departamento de F\'isica, Centro de Nanociencias y Nanotecnolog\'ia, Universidad Nacional Aut\'onoma de M\'exico. Apdo.\ postal 14, 22800, Ensenada, Baja California, M\'exico}

\author{Isaac Soltero}
\affiliation{Escuela Superior de F\'isica y Matem\'aticas, Instituto Polit\'ecnico Nacional. Gustavo A.\, Madero, C.P. 07738, Ciudad de M\'exico, M\'exico}

\author{Francisco Mireles}
\affiliation{Departamento de F\'isica, Centro de Nanociencias y Nanotecnolog\'ia, Universidad Nacional Aut\'onoma de M\'exico. Apdo.\ postal 14, 22800, Ensenada, Baja California, M\'exico}

\date{\today}

\begin{abstract}
Transition-metal dichalcogenide heterostructures exhibit moir\'e patterns that spatially modulate the electronic structure across the material's plane. For certain material pairs, this modulation acts as a potential landscape with deep, trigonally symmetric wells capable of localizing interlayer excitons, forming periodic arrays of quantum emitters. Here, we study these moir\'e localized exciton states and their optical properties. By numerically solving the two-body problem for an interacting electron-hole pair confined by a trigonal potential, we compute the localized exciton spectra for different pairs of materials. We derive optical selection rules for the different families of localized states, each belonging to one of the irreducible representations of the potential's symmetry group $C_{3v}$, and numerically estimate their polarization-resolved absorption spectra. We find that the optical response of localized moir\'e interlayer excitons is dominated by states belonging to the doubly-degenerate $E$ irreducible representation. Our results provide new insights into the optical properties of artificially confined excitons in two-dimensional semiconductors.
\end{abstract}

\maketitle

\section{Introduction}\label{sec:intro}
Atomically thin layers of semiconducting transition-metal dichalcogenides (TMDs) have emerged as a promising optoelectronics platform, based on their valley-dependent optical selection rules\cite{WangYaoPRB_2008,cao2012valley,mak2012control,zeng2012valley} and enhanced optical activity mediated by strongly-bound excitons\cite{MakPRL_MoS2,splendiani2010}. Among these, so-called interlayer excitons (IXs) form in band-mismatched heterobilayers, where electrons and holes preferentially localize in different materials\cite{ceballos2014ultrafast} [Fig.\ \ref{fig:banddiagram}(a)]. Being delocalized across the two crystals, IXs are especially susceptible to the heterostructure's stacking configuration. For instance, it has been shown that the IX energies can be tuned by means of the interlayer twist angle\cite{kunstmann2018momentum}, by their Stark shift in the presence of out-of-plane electric fields\cite{imamoglu2020}, as well as hybridization with bright intralayer exciton states\cite{hXnature2019}. These and other examples\cite{Hsueaax7407,Shimazaki2020,Brotons2020} suggest that manipulating IXs may constitute a viable way of tuning the optical properties of TMD-based heterostructures.

In closely aligned TMD heterobilayers, the slight incommensurability between the two lattices produces a moir\'e pattern\cite{kuwabara1990}: an approximately periodic spatial modulation of the relative shift between the two materials. Due to the long-range periodicity of the moir\'e pattern, different regions of the heterostructure have an approximately commensurate stacking that determines the local electrostatic environment\cite{enaldievPRL,westonNatNano,rosenberger2020}, the value of the heterostructure's band gap\cite{hongyi2017}, and consequently the IX energy. Thus, the moir\'e pattern acts as an approximate superlattice potential for IXs, leading to zone folding and miniband formation\cite{MacDonaldPRL2017,MacDonaldPRB2018}, visible in experiments as a complex fine structure in the material's optical spectrum\cite{hXprb2019,hXnature2019,jin2019,hXreproduced}.

For certain material pairs, such as WSe${}_2$/MoSe${}_2$ and WSe${}_2$/MoS${}_2$, the moir\'e potential exhibits\cite{hongyi2017} deep potential wells capable of localizing IXs\cite{seyler2019,tran2019,gerardot2020}, effectively forming tunable quantum emitter arrays with the periodicity of the moir\'e pattern. The optical response of such states is governed by their symmetry, inherited from that of the potential well and from the carrier Bloch functions at the localization center, giving rise to optical selection rules distinct from those of extended exciton states.

Here, we numerically study the localization and optical spectra of IX states confined by moir\'e potential wells in TMD heterobilayers. We focus on closely aligned, nearly commensurate structures, where the moir\'e supercell length is much larger than the exciton Bohr radius. This allows us to separate the exciton's center of mass (COM) and relative (RM) motions, and solve each problem individually by direct diagonalization methods to compute the low-energy exciton spectrum and wavefunctions. We report a sequence of localized exciton levels identified by their $C_{3v}$ quantum numbers, which is robust for the different material pairs studied in this paper, based on \emph{ab initio} parametrizations of their moir\'e potentials reported by Yu \emph{et al}.\cite{hongyi2017} We give detailed account of these states' optical selection rules for twisted heterobilayers close to parallel ($R$) and anti-parallel ($H$) stacking, and estimate the corresponding absorption spectra by considering the dominant processes mediated by hybridization with intralayer excitons\cite{anomalous_lightcones}. Our analysis reveals that the heterostructure's optical response is dominated by states with orbital wave functions belonging to the $E$ irreducible representation of the group $C_{3v}$. Our results shed new light on the optical properties of IX states in 2D semiconductors confined by artificial potentials.

The rest of this paper is organized as follows: we discuss our theoretical model and its constraints in Sec.\ \ref{sec:model}. In Sections \ref{sec:RM} and \ref{sec:COM} we describe our numerical treatment of the IX relative-motion and center-of-mass problems, respectively, and present calculated spectra for moir\'e localized interlayer excitons in WSe${}_2$/MoSe${}_2$ and WSe${}_2$/MoS${}_2$ heterobilayers. The symmetry properties and optical selection rules of these states are discussed in Sec.\ \ref{sec:optics}, and their numerical absorption spectra are presented in Sec.\ \ref{sec:spectrum}. Our concluding remarks appear in Sec.\ \ref{sec:conclusions}.

\section{Model}\label{sec:model}
\begin{figure}[t!]
\begin{center}
\includegraphics[width=0.9\columnwidth]{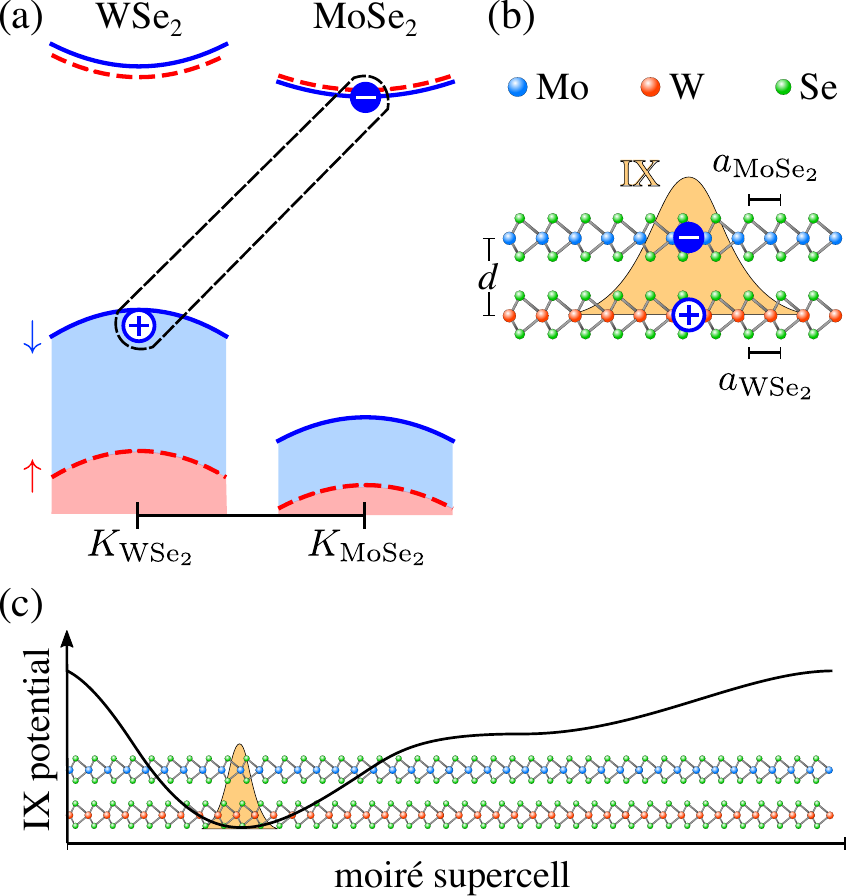}
\caption{(a) $K$-valley band alignment for the type-II heterobilayer WSe${}_2$/MoSe${}_2$. Blue solid (red dashed) curves represent spin-down (spin-up) electronic bands. An electron-hole pair excitation is also depicted, which binds into an IX. (b) Side view of the heterostructure, showing the spatial separation between the electron and hole forming the IX. The relative motion wavefunction is sketched in orange. (c) A broader side view of the heterostructure within a moir\'e supercell. The moir\'e potential is shown, with an IX localized at its minimum.}
\label{fig:banddiagram}
\end{center}
\end{figure}

Our focus will be on semiconducting TMD heterobilayers with staggered band alignment, whose conduction- and valence-band edge states are localized in different layers. We use the nomenclature MX${}_2$/M$'$X$'{}_2$, where M$'$X$'{}_2$ (MX${}_2$) is the TMD layer to which the bottom of the conduction band (top of the valence band) belongs,  as sketched in Fig.\ \ref{fig:banddiagram}(a). To describe interlayer excitons in the heterostructure, formed by an M$'$X$'{}_2$ electron of effective mass $m_{\rm e'}$ and an MX${}_2$ hole of effective mass $m_{\rm h}$, we use the two-body Hamiltonian
\begin{equation}\label{eq:Heh}
	H_{\rm e-h}=\frac{P^2}{2M} + \frac{p^2}{2\mu} + U_{\rm K}(|\rr_{\rm e}-\rr_{\rm h}|) + V_{\rm M}(\rr_{\rm e},\rr_{\rm h}).
\end{equation}
Here, $\PP$ and $\pp$ represent the total and relative momenta;
\begin{equation}
M = m_{\rm e'}+m_{\rm h},\quad \mu = \frac{m_{\rm e'}m_{\rm h}}{m_{\rm e'}+m_{\rm h}},
\end{equation}
are the total and reduced masses of the electron-hole system; $U_{\rm K}$ is the electrostatic interaction between the electron and hole at positions $\rr_{\rm e}$ and $\rr_{\rm h}$; and $V_{\rm M}$ represents the periodic moir\'e potential.

We consider electrostatic interactions at the semiclassical and macroscopic level, which appropriately describe the exciton spectra of two-dimensional TMDs according to experiments\cite{chernikov2014}. We note in passing that this approximation is not essential for the numerical methods used below. The heterostructure is assumed to be embedded in an environment with average dielectric constant $\epsilon$, which together with the in-plane electric susceptibilities $\kappa$ and $\kappa'$ of the two layers defines the length scales (Table \ref{tab:parameters}) $r_*=2\pi\kappa/\epsilon$ and $r_*'=2\pi\kappa'/\epsilon$, below which electrostatic interactions in the corresponding TMD layer are screened. For simplicity, we adopt the long-range approximation defined by $\rho\equiv|\rr_{\rm e}-\rr_{\rm h}|\gg r_*,\,r_*'$, which gives the electron-hole interaction in the Keldysh form (see Appendix \ref{app:interaction})\cite{danovich_2018,keldysh}
\begin{equation}\label{eq:UK}
	U_{\rm K}(\rho) = \frac{-e^2\pi}{ 2 \epsilon r_{\rm eff}}\left[H_0\left(\frac{\rho}{r_{\rm eff}}\right) - Y_0\left(\frac{\rho}{r_{\rm eff}}\right) \right],
\end{equation}
where $H_0$ and $Y_0$ are a Struve function and a Bessel function of the second kind, respectively, and $e$ is the elementary charge. The effective screening length $r_{\rm eff}=r_*+r_*'+d$ accounts for the dielectric response of the additional layer as well as the out-of-plane separation between their carriers, $d$ [Fig.\ \ref{fig:banddiagram}(b)]. For intralayer excitons in the heterostructure, the appropriate definition is $r_{\rm eff}=r_*+r_*'$, whereas in a monolayer $r_{\rm eff}=r_*$. As discussed in Ref.\ \onlinecite{danovich_2018}, this approximation overestimates the short-range interactions in the case of IXs, and the binding energies (Bohr radii) computed from \eqref{eq:UK} must be interpreted as an upper (lower) limit.
\begin{table}[b!]
\caption{In-plane lattice constants $a$, effective electron and hole masses $m_{\rm e}$ and $m_{\rm h}$, and screening lengths $r_*$ \emph{in vacuo} ($r_* = 2\pi\kappa/\epsilon_{\rm vac}$, $\epsilon_{\rm vac}=1$) for the main semiconductor TMDs, extracted from the experimental and \emph{ab initio} literature. We also show the (maximum) moir\'e superlattice parameters $a_{\rm M}$ for different heterobilayers, assuming perfect rotational alignment. }
\begin{center}
\begin{tabular}{l r@{.}l   r@{.}l r@{.}l   r@{.}l    |  l@{/}l  r@{.}l  r@{.}l}
\hline\hline
TMD & \multicolumn{2}{c}{$a$ [\AA]}  & \multicolumn{2}{c}{$m_{\rm e}/m_{0}$}  & \multicolumn{2}{c}{$m_{\rm h}/m_0$} & \multicolumn{2}{c|}{$r_*$ [\AA]} & \multicolumn{2}{c}{Heterostructure}  & \multicolumn{2}{c}{$d$ [\AA]} & \multicolumn{2}{c}{$a_{\rm M}$ [nm]} \\
\hline
MoS${}_2$   & 3&160${}^{\rm a}$             & 0 & 70${}^{\rm d}$       &0&70${}^{\rm h}$        &38&62${}^{\rm i}$     & MoSe${}_2$&MoS${}_2$ &      6&97${}^{\rm j}$  & 7&21 \\
                     & 3&140${}^{\rm b}$             &\multicolumn{2}{c}{\,}   &\multicolumn{2}{c}{\,}   &\multicolumn{2}{c|}{\,}     & MoSe${}_2$&WS${}_2$ &    6&88${}^{\rm j}$  & 7&42 \\
MoSe${}_2$ & 3&299${}^{\rm b}$             & 0&80${}^{\rm e}$        &0&50${}^{\rm h}$         &39&79${}^{\rm i}$     & WSe${}_2$&MoS${}_2$ &           6&83${}^{\rm j}$  & 7&72  \\
                     & 3&288${}^{\rm a}$             &\multicolumn{2}{c}{\,}   &\multicolumn{2}{c}{\,}   &\multicolumn{2}{c|}{\,}      & WSe${}_2$&MoSe${}_2$&   7&02${}^{\rm j}$  & 108&18  \\
WS${}_2$    & 3&154${}^{\rm a,c}$           &0&27${}^{\rm f}$          &0&50${}^{\rm h}$         &37&89${}^{\rm i}$      & WSe${}_2$&WS${}_2$&              6&86${}^{\rm j}$  &  7&97  \\
WSe${}_2$  & 3&286${}^{\rm a}$              &0&50${}^{\rm g}$         &0&45${}^{\rm i}$          &45&11${}^{\rm i}$       & WS${}_2$&MoS${}_2$ &             6&76${}^{\rm j}$   &  248&38  \\
                    & 3&282${}^{\rm b}$              &\multicolumn{2}{c}{\,}   &\multicolumn{2}{c}{\,}   &\multicolumn{2}{c|}{\,}\\
                    \hline\hline
\end{tabular}\\
${}^{\rm a}$Reference [\onlinecite{wilson1969transition}], ${}^{\rm b}$Reference [\onlinecite{al1972preparation}], ${}^{\rm c}$Reference [\onlinecite{yang1996li}], ${}^{\rm d}$Reference [\onlinecite{MoS2mass}], ${}^{\rm e}$Reference [\onlinecite{MoSe2mass}], ${}^{\rm f}$Reference [\onlinecite{kormanyos}], ${}^{\rm g}$Reference [\onlinecite{WSe2mass}], ${}^{\rm h}$Reference [\onlinecite{ArpesMasses}], ${}^{\rm i}$Reference [\onlinecite{Fallahazad2016}], ${}^{\rm i}$Reference [\onlinecite{mostaani2017}], ${}^{\rm j}$Reference [\onlinecite{xu2018role}]
\end{center}
\label{tab:parameters}
\end{table}%

The moir\'e potential $V_{\rm M}$ varies over length scales of the order of the moir\'e superlattice constant [Fig.\ \ref{fig:banddiagram}(c)]
\begin{equation}\label{eq:aM}
	a_{\rm M} = \frac{a_{<}}{\sqrt{\delta^2+\theta^2}},
\end{equation}
defined by the interlayer twist angle $\theta \ll 1$ and lattice mismatch $\delta=1-a_</a_>$, where $a_{>}$ ($a_<$) is the larger (smaller) lattice constant of the two TMD layers. Based on Eq.\ \eqref{eq:aM} and the known lattice constants of the four main semiconductor TMDs (Table \ref{tab:parameters}), we estimate $a_{\rm M}$ for both $R$ and $H$ structures to be of order $10\,{\rm nm}$ for heterobilayers with different chalcogens, and $100\,{\rm nm}$ for those with matching chalcogens, like WSe${}_2$/MoSe${}_2$ and WS${}_2$/MoS${}_2$. By contrast, $U_{\rm K}$ binds electrons and holes into excitons with Bohr radii\cite{berkelbach2013,danovich_2018,hXprb2019} $a_{\rm B} \approx 10$-$20\,\text{\AA}$, over which $V_{\rm M}$ varies slowly. This allows us to treat the electron-hole pair as point-like, and located at the COM position $\RR$. The moir\'e potential then becomes
\begin{equation}\label{eq:decouple}
	V_{\rm M}(\rr_{\rm e},\rr_{\rm h}) \approx V_{\rm M}(\RR),\, \RR = \frac{m_{\rm e}\rr_{\rm e} + m_{\rm h}\rr_{\rm h}}{M},
\end{equation}
neglecting any RM dependence and making the Hamiltonian \eqref{eq:Heh} separable into a COM part and a RM part. This approach is especially well suited for studying excitons in chalcogen-matched structures such as WSe${}_2$/MoSe${}_2$ and WS${}_2$/MoS${}_2$ at small twist angles, where the close match between lattice constants guarantees a large moir\'e periodicity and thus the scale separation described above. Conversely, we expect this approach to be as well suited for the case of strong misalignment angles and for heterobilayers formed with TMDs containing different chalcogens. In the following we shall focus on the former case, setting the twist angle to zero.

With \eqref{eq:decouple}, solutions to the Hamiltonian \eqref{eq:Heh} have the form
\begin{equation}\label{eq:psi}
\Psi(\rr_{\rm e},\,\rr_{\rm h}) = F(\RR)f(\rrho),
\end{equation}
where $\rrho\equiv\rr_{\rm e}-\rr_{\rm h}$. Since the Keldysh potential is isotropic in the plane it preserves angular momentum, and the second factor in \eqref{eq:psi} can be written as
\begin{equation}\label{eq:fmn}
	f(\rrho) \equiv f_m(\rho,\phi) = \frac{\exp{im\phi}}{\sqrt{2\pi}}\chi_m(\rho),
\end{equation}
where the integer $m$ is the RM angular momentum quantum number, $\rho\equiv |\rrho|$ and $\phi$ is the azimuthal angle of vector $\rrho$. Operating on \eqref{eq:psi} with $H_{\rm e-h}$ and dividing by the same wavefunction yields the two eigenvalue problems
\begin{subequations}
\begin{equation}\label{eq:RM}
	\left[-\left(\frac{\partial^2}{\partial \rho^2} + \frac{1}{\rho}\frac{\partial}{\partial \rho} - \frac{m^2}{\rho^2}\right) + \frac{2\mu}{\hbar^2}\left( U_{\rm K}(\rho) - E_f^m \right)   \right]f(\rrho) = 0,
\end{equation}
\begin{equation}\label{eq:COM}
	\left[-\nabla_{\RR}^2 + \frac{2M}{\hbar^2}\left( V_{\rm M}(\RR) - E_F \right)  \right]F(\RR) = 0,
\end{equation}
\end{subequations}
with the total energy of state $\Psi$ given by $E=E_F+E_f^m$.

\subsection{The relative motion problem}\label{sec:RM}
\begin{figure}[h!]
\begin{center}
\includegraphics[width=\columnwidth]{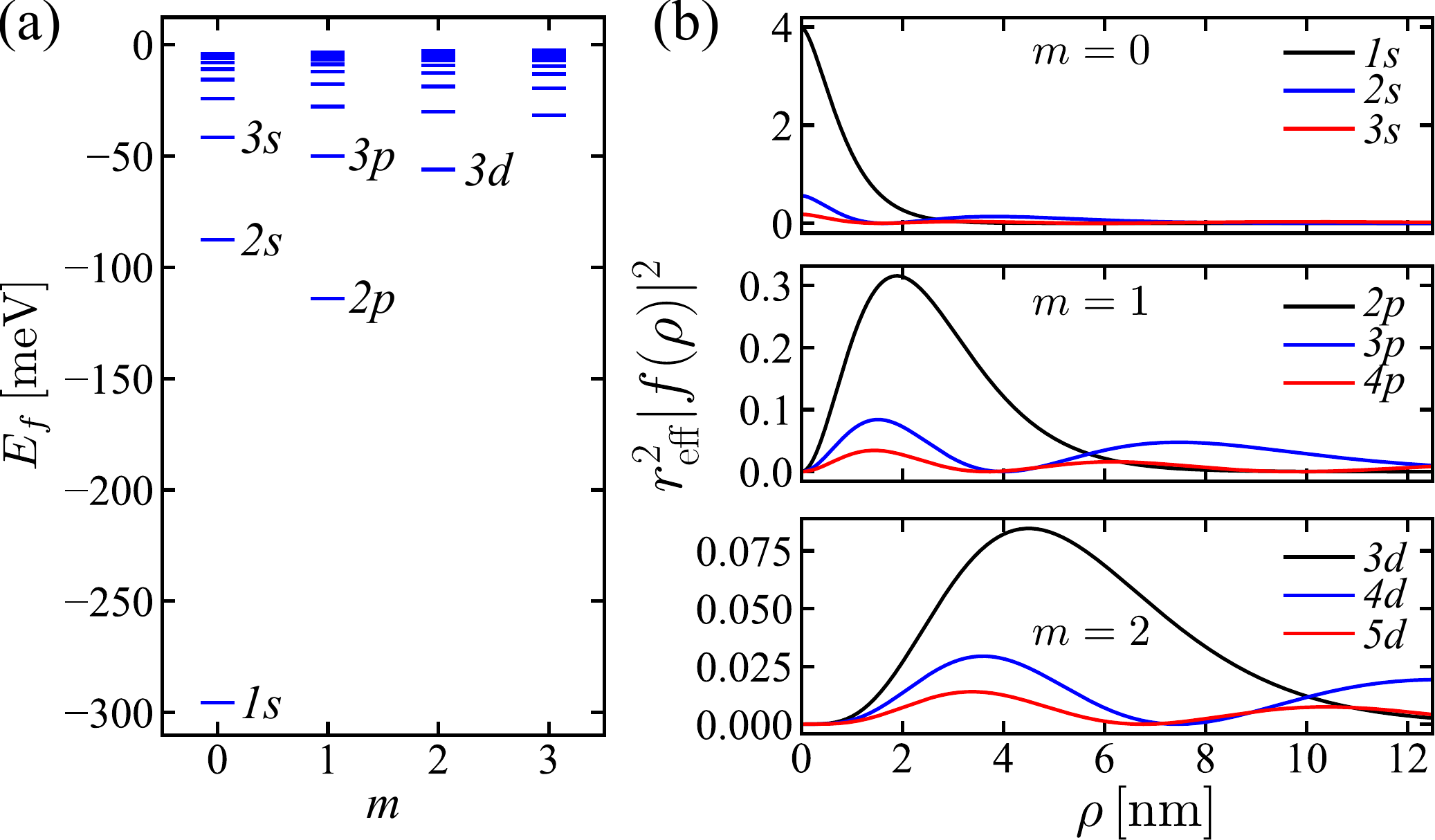}
\caption{(a) Binding energies of intralayer excitons in WS${}_2$ on a SiO${}_2$ substrate, as obtained by the direct diagonalization method described in the text. These results are in good agreement with the experiments of Ref.\ \onlinecite{chernikov2014}. (b) The radial relative-motion wavefunctions corresponding to some of the states shown in (a).}
\label{fig:WS2X}
\end{center}
\end{figure}

Equation \eqref{eq:RM} represents a two-dimensional hydrogenic problem with a non-trivial interaction whose form is logarithmic in the short-range limit\cite{2DHydrogen1}, and of Coulomb type at long distances\cite{2DHydrogen2}. Different approaches have been taken to solve this equation in the context of exciton formation in TMDs, including variational methods\cite{berkelbach2013}, quantum Monte Carlo simulations\cite{mostaani2017,danovich_2018}, and finite elements calculations\cite{danovich_2018}. Here, we adopt an economical numerical method based on direct diagonalization in a truncated basis, first introduced in nuclear physics\cite{griffin1957}, and more recently used in the context of acceptor states in semiconductors\cite{baldereschi1973,mireles_1998,mireles_1999}. The basis functions
\begin{equation}\label{eq:basis}
	\chi_j^m (\rho) \equiv (\beta \rho)^{|m|} \exp{-\beta_j \rho},\, 1\le j\le N,
\end{equation}
are inspired by analytical solutions to the 2D hydrogen atom problem\cite{2DHydrogen2}, and share their general behavior at $\rho=0$ and $\rho\rightarrow \infty$. Each basis element is defined by its decay length $\beta_j^{-1}$, which varies discretely between two values $\beta_N^{-1}$ and $\beta_1^{-1}$. An arbitrary length scale $\beta^{-1}$ has also been introduced, to make each function dimensionless. We have chosen $ \beta_N^{-1} \lesssim r_{\rm eff}  \ll \beta_1^{-1}$ to cover the entire range of Bohr radii that the low-energy excitons are likely to take. The values of $\beta_\ell$ are spaced logarithmically as
\begin{equation}
	\beta_j = \beta_1\exp{\xi(j-1)},\quad \xi = (N-1)^{-1}\log{\left( \beta_N/\beta_1 \right)},
\end{equation}
to cover more densely length scales of the order of the screening lengths and below. Then, the radial part of the RM wavefunction can be expanded in this basis as
\begin{equation}\label{eq:radialsol}
	\chi_m(\rho) = \sum_{j=1}^NA_j^{m}\chi_j^m(\rho).
\end{equation}
Substituting \eqref{eq:radialsol} into \eqref{eq:RM}, left-multiplying by $\chi_m(\rho)$ and integrating leads to the generalized eigenvalue problem
\begin{equation}\label{eq:geneig}
	\left[ H^m - E_fS^m \right]A^m = 0,
\end{equation}
where we have defined the column vector $A^m=(A_1^m,\ldots, A_N^m)^T$, and the matrices $H^m$ and $S^m$ are shown explicitly in Appendix \ref{app:matrices}. Equation \eqref{eq:geneig} is solved numerically to obtain a set of eigenvalues $\{ E_f^{m,n} \}$ and eigenvectors $\{A^{m,n}\}$, where $n$ is the principal quantum number.

To test this method, we have evaluated the intralayer exciton spectrum of WS${}_2$ on a SiO${}_2$ substrate by setting $\varepsilon=(\varepsilon_{\rm vacuum}+\varepsilon_{\rm SiO_2})/2=2.4$, and $r_{\rm eff}=2\pi\kappa_{\rm WS_2}/\varepsilon=15.79\,\text{\AA}$ in Eq.\ \eqref{eq:UK}. The first few energies and radial wavefunctions are presented in Fig.\ \ref{fig:WS2X}. The obtained values reproduce the non-Rydberg sequence reported experimentally in Ref.\ \onlinecite{ziliang2014} and theoretically in Ref.\ \onlinecite{vanderdonck2019}. Moreover, the energy spacings between states $1s$, $2s$ and $3s$ are a good match\footnote{As mentioned in Sec.\ \ref{sec:model}, short-ranged interlayer interactions are overestimated by the Keldysh formula \eqref{eq:UK}. This results in overestimation of the binding energies of the lowest energy states, particularly $1s$.} to those reported in Ref.\ \onlinecite{chernikov2014}.

Table \ref{tab:resinter} shows our results for the low-energy interlayer excitons in all TMD heterobilayers formed with Mo, W, S and Se, also between a SiO${}_2$ substrate and air/vacuum. The band alignment of each structure, which determines to which material the electron and hole making up the IX belong, has been taken from \emph{ab initio} calculations\cite{gong2013band}.

\begin{table}[h!]
\caption{Calculated binding energies of the lowest-lying interlayer excitons for different semiconducting TMD heterobilayers on a typical SiO${}_2$ substrate. All heterostructures shown have type-II (staggered) band gaps\cite{gong2013band}. The electron and hole masses are taken from Table \ref{tab:parameters}, according to which layer contains the highest valence-band edge and the lowest conduction-band edge, as reported in Ref.\ \onlinecite{gong2013band}.}
\begin{center}
\begin{tabular}{l@{/}l   c  c  c  c  c  c}
\hline\hline
\multicolumn{2}{c}{\,}                                                              & \multicolumn{6}{c}{Binding energy [meV]}\\
\multicolumn{2}{c}{\raisebox{1.5ex}[0pt]{Heterostructure}} & $1s$ & $2p$ & $2s$  & $3d$  & $3p$ & $3s$\\
\hline
 MoSe${}_2$ & MoS${}_2$ &       195 & 100 &  79 & 61 & 53 & 45      \\ %
 MoSe${}_2$ & WS${}_2$ &         167 &  78  &  61& 44 & 39 & 32       \\ %
 WSe${}_2$ & MoS${}_2$  &       185  &  94  & 75 & 57 & 50 & 42     \\ %
 WSe${}_2$ & MoSe${}_2$ &      185  &  96  & 76 & 59 & 51 & 43      \\ %
 WSe${}_2$ & WS${}_2$     &      159 &  75  & 58 & 42 & 37 & 31        \\ %
 WS${}_2$ & MoS${}_2$      &     199  &101  & 80 & 61 & 54 & 45       \\ %
\hline\hline
\end{tabular}
\end{center}
\label{tab:resinter}
\end{table}%

\begin{figure}[t]
\begin{center}
\includegraphics[width=\columnwidth]{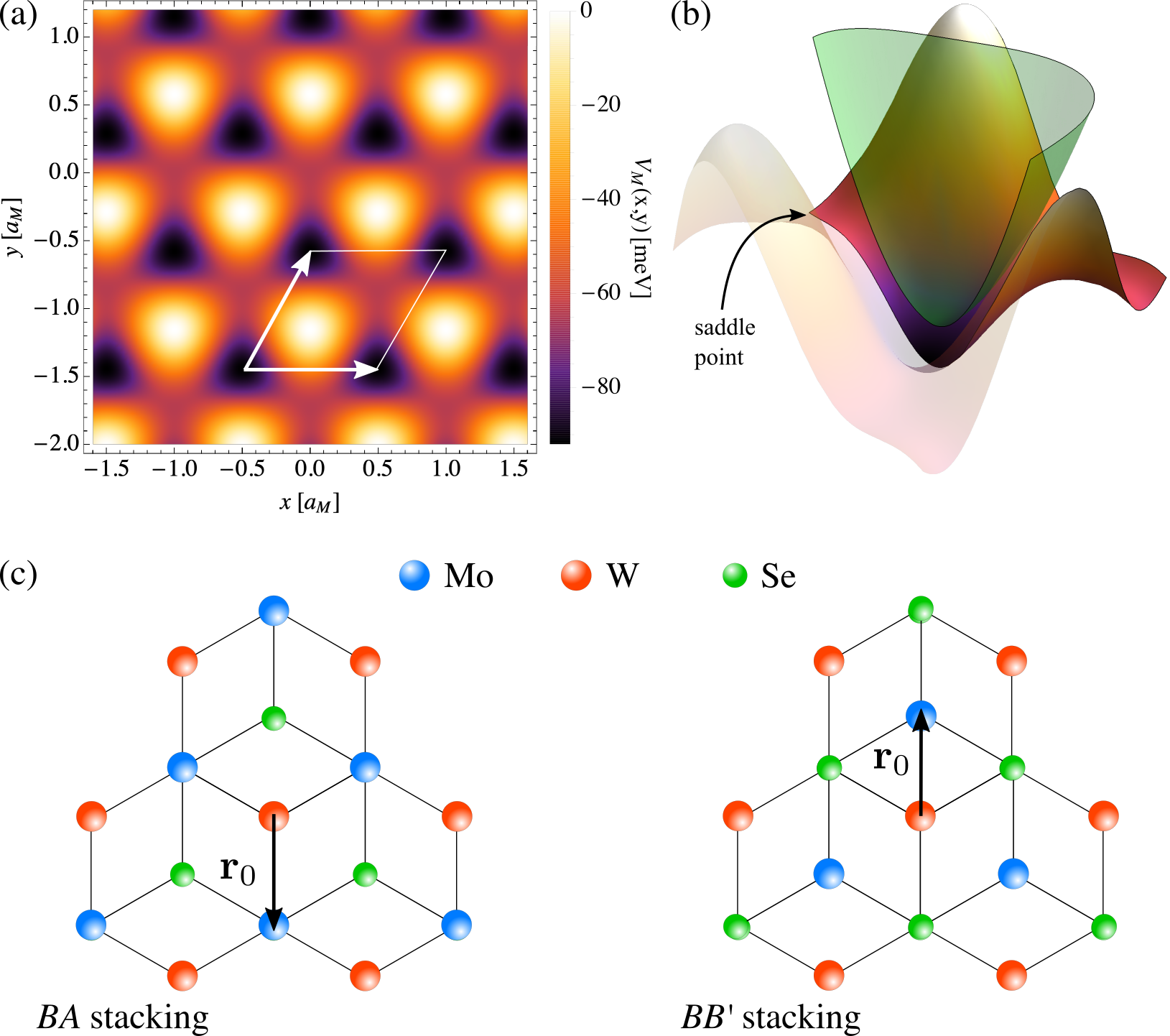}
\caption{(a) \emph{Ab initio} moir\'e potential for interlayer excitons in a fully aligned, $R$-stacked WSe${}_2$/MoSe${}_2$ heterobilayer, as calculated by Yu \emph{et al}.\cite{hongyi2017} The moir\'e supercell is shown in white, with the two primitive moir\'e vectors represented by white arrows. (b) Potential profile around one of the minima, resembling a $C_{3v}$-symmetric harmonic potential below the saddle point energy. Our fitting of this potential well with Eq.\ \eqref{eq:fit} is shown in green, offset out of the plane for clarity. (c) Top view of the local stacking configurations at the moir\'e potential minima for $R$-stacked ($BA$) and $H$-stacked ($BB'$) TMD heterobilayers. In each case, the shortest in-plane vector $\mathbf{r}_0$ joining the W atom to the M atom is shown.}
\label{fig:MoireWSe2MoSe2}
\end{center}
\end{figure}

\subsection{The center-of-mass motion problem}\label{sec:COM}
The moir\'e potential $V_{\rm M}$ appearing in Eq.\ \eqref{eq:COM} has been modeled and parametrized by Yu \emph{et al.} in Ref.\ \onlinecite{hongyi2017}, based on \emph{ab initio} results for the band-gap and band alignment variation across different TMD heterobilayers. While this and other potentials found in the \emph{ab initio} literature\cite{MacDonaldPRL2017,MacDonaldPRB2018} ignore recently discovered lattice relaxation phenomena\cite{enaldievPRL,westonNatNano,rosenberger2020} that lead to domain formation and piezoelectric effects, we shall show below that the potentials of Yu \emph{et al.}\ produce localized IX spectra that compare favorably with available experimental data\cite{seyler2019}. Fig.\ \ref{fig:MoireWSe2MoSe2}(a) shows the superlattice potential for interlayer excitons in one such case: $R$-stacked WSe${}_2$/MoSe${}_2$, where at zero twist angle the moir\'e lattice constant is about $100\,{\rm nm}$. The potential landscape contains periodic potential wells interconnected by saddle points and surrounded by three maxima. Each of these wells is $C_{3v}$ symmetric about its minimum, and for energies below the saddle point they can be modeled by a simple trigonally-warped harmonic potential of the form
\begin{equation}\label{eq:fit}
	\tilde{V}_{\rm M}(R,\Phi) = V_{0} +  \frac{M\omega^2}{2}R^2\left[1-\delta\cos{(3\Phi + \varphi)} \right],
\end{equation}
centered at the well minimum, where $R$ and $\Phi$ are the magnitude and polar angle of the COM vector $\RR$, $R_0=\sqrt{\hbar M^{-1}\omega^{-1}}$ is the harmonic oscillator characteristic length, and $\omega$, $\delta$ and $\varphi$ are fitting parameters. This is exemplified in Fig.\ \ref{fig:MoireWSe2MoSe2}(b). The corresponding fitting parameters for interlayer excitons in this and other TMD heterobilayers are reported in Table \ref{tab:fit}.
\begin{table}[h!]
\caption{Fitting parameters for the model Eq.\ \eqref{eq:fit}, corresponding to the wells appearing in the moir\'e potentials reported in Ref.\ \onlinecite{hongyi2017}, based on DFT calculations. The localized state width $R_0$ and the barrier energy $E_*$ are also reported.}
\begin{center}
\begin{tabular}{c| r@{.}l r@{.}l  r@{.}l c r@{/}l}
\hline\hline
\textbf{Heterostructure} & \multicolumn{2}{c}{$\hbar\omega$ [meV]} & \multicolumn{2}{c}{$\delta$} & \multicolumn{2}{c}{$R_0$ [\AA]} & $E_*$ [meV] & \multicolumn{2}{c}{$\varphi$}\\ 
\hline
$R$-WSe${}_2$/MoSe${}_2$    & 2&12 & 0&290  & 53&64  & 28 & $\pi$&$2$ \\  
$H$-WSe${}_2$/MoSe${}_2$    & 1&23 & 0&292  & 70&52  & 14 & $-\pi$&$2$ \\ 
$R$-WSe${}_2$/MoS${}_2$      & 54&83 &  0&330 & 10&99 & 28 & $-\pi$&$2$  \\ 
$H$-WSe${}_2$/MoS${}_2$      & 30&88 &  0&019 & 14&65 & 58 & $-\pi$&$2$ \\
\hline\hline
\end{tabular}
\end{center}
\label{tab:fit}
\end{table}%

Unlike the true potential $V_{\rm M}$, Eq.\ \eqref{eq:fit} is unbounded and will always produce localized states, the lowest of which will appear at an energy $E_{F,1} \approx \hbar \omega$ above the potential bottom. Nonetheless, as long as the energy eigenvalue $E_{F,\ell}$ is well below the saddle point energy $E_*$ [see Fig.\ \ref{fig:MoireWSe2MoSe2}(b)] and the wavefunction decays rapidly enough, the corresponding state will be a good approximation to a localized state of the finite potential well. This is, indeed, the case for $R$-WSe${}_2$/MoSe${}_2$, where $\hbar \omega=2\,{\rm meV}$ is much lower than the saddle point barrier, and the state width, defined by $R_0$, is a full order of magnitude shorter than the moir\'e superlattice constant (Table \ref{tab:fit}). This suggests that the moir\'e localized IXs recently reported experimentally by Seyler \emph{et al.}\cite{seyler2019} and Tran \emph{et al.}\cite{tran2019} are well described by the potential \eqref{eq:fit}. Our estimated values for $\hbar \omega$ are, in fact, consistent with the shortest energy differences between moir\'e localized IX resonances reported in Ref.\ \onlinecite{seyler2019} for both $R$- and $H$-stacked WSe${}_2$/MoSe${}_2$, indicating that the potential parametrization by Yu \emph{et al.}\cite{hongyi2017} correctly describes the localization physics.  By contrast, the parameters reported in Table \ref{tab:fit} indicate that chalcogen-mismatched heterostructures like WSe${}_2$/MoS${}_2$ may not be well described by this approach, since their moir\'e periodicities are comparable to the localized state width $R_0$. A notable case is $R$-WSe${}_2$/MoS${}_2$, for which $\hbar \omega$ greatly exceeds the saddle-point energy, such that localized IXs are not to be expected in this heterostructure.

To find the energies of IXs localized by the potential \eqref{eq:fit}, we numerically solve the eigenvalue problem \eqref{eq:COM} by direct diagonalization over the basis of eigenstates of the 2D harmonic oscillator (2DHO), expressed in cylindrical coordinates as
\begin{equation}\label{eq:basisCOM}
\begin{split}
	\psi_{j,\mm}(R,\Phi)=\frac{\exp{i\mm\Phi}}{\sqrt{2\pi N_{j,\mm}}}\left(\frac{R}{R_0} \right)^{|\mm|}\exp{-R^2/2R_0^2}L_{\tfrac{j-|\mm|}{2}}^{|\mm|}(R^2/R_0^2).
\end{split}
\end{equation}
Here, $L_{n}^{m}$ are associated Laguerre polynomials, and the normalization factors are
\begin{equation}
	N_{j,\mm} = \frac{R_0^2}{2}\left(\frac{j-|\mm|}{2} +1 \right)_{|\mm|},
\end{equation}
with $(x)_n=\Gamma(x+n)/\Gamma(x)$ the Pochhammer symbol. The principal quantum number $j$ is a positive integer or zero, and determines the state energy as $E_{j}^0=\hbar\omega (j + 1)$. The angular momentum quantum number $\mm$ is restricted such that $|\mm|\le j$ and $j-|\mm|$ is an even number, giving a total degeneracy of $j+1$ for the basis state $\psi_{j,\mm}$.

Unlike the case of electron-hole interactions, the $C_{3v}$-symmetric potential $\tilde{V}_{\rm M}$ only preserves angular momentum $\mm$ modulo 3, and matrix elements between states with different angular momenta must be considered. The appropriate quantum number is then
\begin{equation}
	\bar{\mm}= \mm\bmod 3,
\end{equation}
taking the values $-1,\,0$ and $1$ for states belonging to different irreducible representations (irreps) of group $C_{3v}$.

Substituting $\psi_{j,\mm}$ into Eq.\ \eqref{eq:COM}, left-multiplying by $\psi_{j',\mm'}$ and integrating, we get the eigenvalue problem
\begin{equation}\label{eq:eigenCOM}
\begin{split}
	&[E_F-\hbar \omega (j+1)]\delta_{j',j}\delta_{\mm',\mm}\\ &+ \frac{\delta}{4}\frac{\hbar \omega\,V_{j',j}^{\mm',\mm}\delta_{|\mm'-\mm|,3}}{\sqrt{\left(\tfrac{j'-|\mm'|}{2} +1\right)_{|\mm'|}\left(\tfrac{j-|\mm|}{2}+1 \right)_{|\mm|} }} = 0.
\end{split}
\end{equation}
Explicit formulae for $V_{j',j}^{\mm',\mm}$ are provided in Appendix \ref{app:melems}. Equation \eqref{eq:eigenCOM} may be divided into three independent blocks with different $\bar{\mm}$ eigenvalue, each of which can be solved numerically by truncating the basis at principal quantum number $j_{\rm max}\sim 10$, determined by convergence of the low-lying energy eigenvalues. Such convergence is guaranteed, as it can be numerically shown that for any given $\mm'=\mm\pm3$, $V_{\mm',\mm}^{j',j}$ decays rapidly with $|j'-j|$.

\begin{figure}[t!]
\begin{center}
\includegraphics[width=\columnwidth]{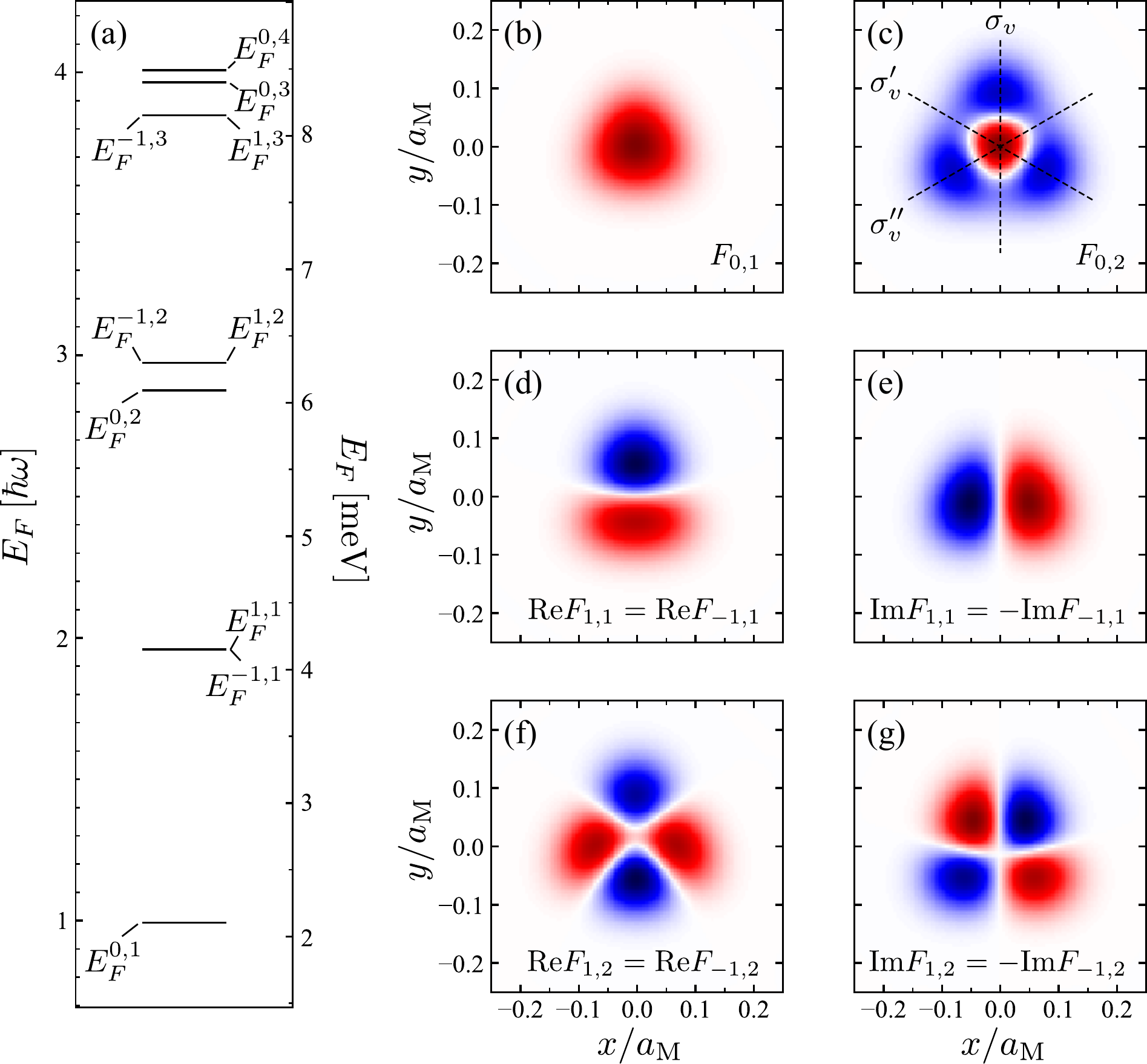}
\caption{(a) Spectrum of moir\'e bound states in $R$-WSe${}_2$/MoSe${}_2$, shown both in units of $\hbar\omega$ and in ${\rm meV}$, according to the parameters reported in Table \ref{tab:fit}. (b)-(g) Wave functions of the six lowest energy states, each normalized to its maximum amplitude for illustration purposes. The $x$ axis has been chosen to cancel out the angle $\varphi$ in Eq.\ \eqref{eq:fit}. Blue (red) indicates positive (negative) values, whereas white represents a zero value. According to their properties under $C_3$ rotations and mirror reflections about the axes shown in panel (c), modes $F_{0,0}$ and $F_{0,1}$ are associated to the $A_1$ irrep, and $F_{\pm1,0}$ and $F_{\pm1,1}$ to the $E$ irrep of the point group $C_{3v}$ of the potential \eqref{eq:fit}.}
\label{fig:RWSe2MoSe2wavefunction}
\end{center}
\end{figure}

Figure \ref{fig:RWSe2MoSe2wavefunction}(a) shows the low-energy spectrum $\{E_F^{\bar{\mm},\ell}\}$ of moir\'e bound interlayer excitons in $R$-WSe${}_2$/MoSe${}_2$, as well as the wavefunctions $F_{\bar{\mm},\ell}$ for the first six states. The corresponding energies for this and other TMD heterostructures are listed in Table \ref{tab:COMlevels}. For simplicity, in both cases we take the potential bottom $V_0$ as the energy reference.  As may be expected by analogy with the isotropic case (2DHO), the ground state, labeled $F_{0,1}$, originates from block $\bar{\mm}=0$. This state can be viewed as the 2DHO state $\psi_{0,0}$, weakly modified by second-order perturbations from states $\psi_{j\ge3,\mm\pm3}$, the lowest of which appear $3\hbar\omega$ higher in energy.  The perturbative parameter is then $\le\delta/12\sim10^{-2}$, and  $E_F^{0,0}\approx \hbar \omega$ to within less than $0.1\%$. The situation is qualitatively similar for the next two energy levels, a degenerate doublet formed by a state from block $\bar{\mm}=1$ and one from block $\bar{\mm}=-1$, with energies $E_F^{\pm1,1}\approx 2\hbar\omega$ to within $5\%$ accuracy. Finally, the 2DHO degenerate triplet formed by $\psi_{2,-2}$, $\psi_{2,0}$ and $\psi_{2,2}$ is split by trigonal warping of the potential into a lower singlet $F_{0,2}$ and a higher degenerate doublet $F_{\pm1,2}$, separated by a gap of $\approx 0.2\hbar \omega$ or $0.5\,{\rm meV}$. Similar level splittings appear for the entire spectrum from this point on.  By comparison, Ref.\ \onlinecite{seyler2019} reports IX photoluminescence (PL) peaks as narrow as $100\,\mu{\rm eV}$, indicating that experimental observation of the predicted broken degeneracies is currently possible. The same sequence of quantum numbers is found also for $H$-WSe${}_2$/MoSe${}_2$ and $H$-WSe${}_2$/MoS${}_2$, as reported in Table \ref{tab:COMlevels}.
\begin{table}[h!]
\caption{Calculated energies of the lowest-lying moir\'e localized IXs in different semiconducting TMD heterobilayers. All energies are measured with respect to the corresponding potential well bottom $V_0$ in Eq.\ \eqref{eq:fit}. For $H$-WSe${}_2$/MoS${}_2$ only the first energy is below the saddle-point barrier.}
\begin{center}
\begin{tabular}{l@{/}l   r@{.}l  r@{.}l  r@{.}l  r@{.}l  r@{.}l  r@{.}l}
\hline\hline
\multicolumn{2}{c}{\,}                                                              & \multicolumn{12}{c}{Localized IX energy [meV]}\\
\multicolumn{2}{c}{\raisebox{1.5ex}[0pt]{Heterostructure}} & \multicolumn{2}{c}{$E_F^{0,1}$} & \multicolumn{2}{c}{$E_F^{\pm1,1}$} & \multicolumn{2}{c}{$E_F^{0,2}$}  & \multicolumn{2}{c}{$E_F^{\pm1,2}$}  & \multicolumn{2}{c}{$E_F^{\pm1,3}$} & \multicolumn{2}{c}{$E_F^{0,3}$}\\
\hline
 $R$-WSe${}_2$ & MoSe${}_2$ &       2&105 &   4&153 &    6&095 &   6&303 &     8&158 &     8&406      \\ %
 $H$-WSe${}_2$ & MoSe${}_2$ &       1&221 &   2&410 &    3&535 &   3&656 &     4&731 &     4&876      \\ %
 $H$-WSe${}_2$ & MoS${}_2$   &     30&877 & 61&754 &  92&621 & 92&637 & 123&498 & 123&517      \\ %
\hline\hline
\end{tabular}
\end{center}
\label{tab:COMlevels}
\end{table}%

\begin{figure*}[t!]
\begin{center}
\includegraphics[width=1.6\columnwidth]{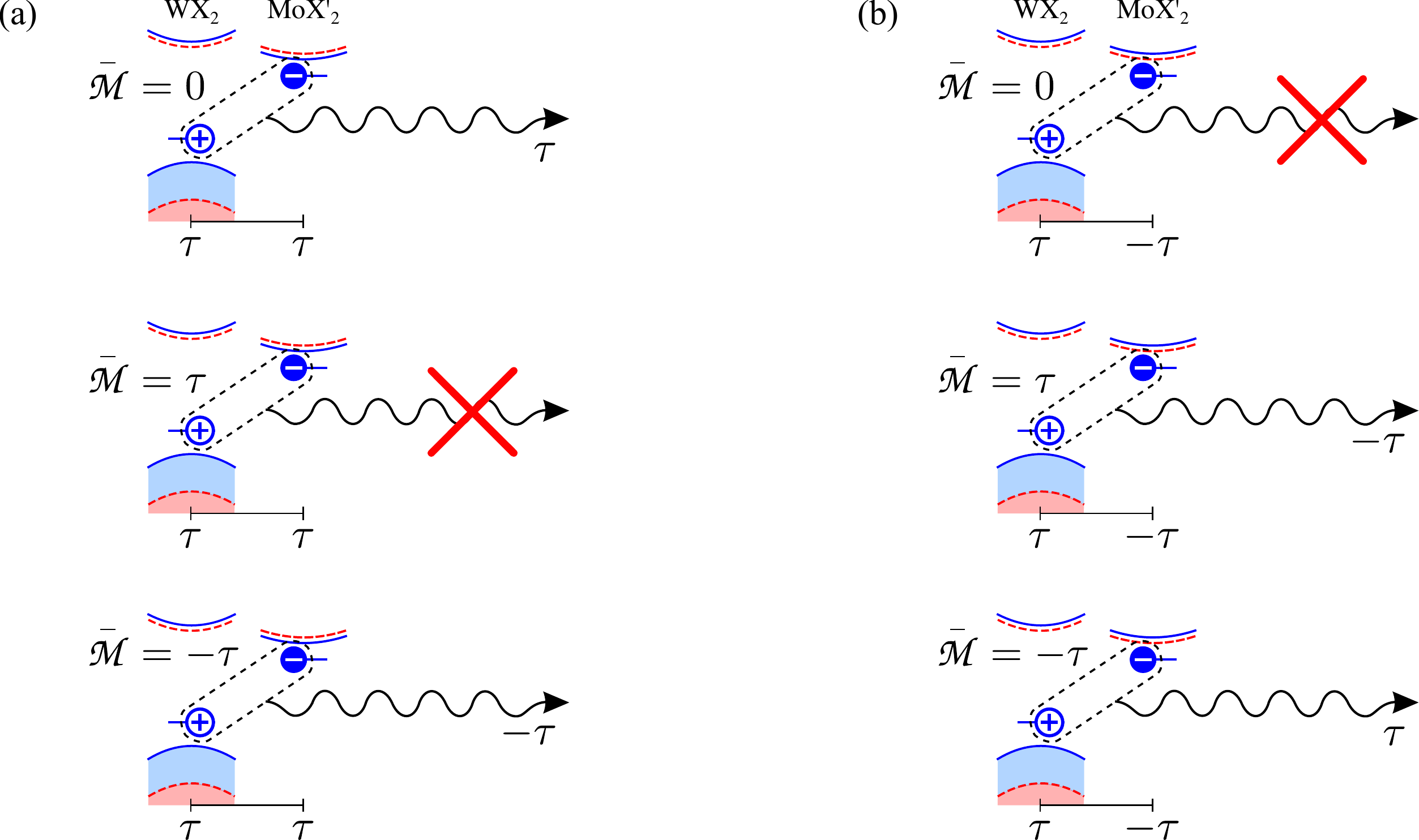}
\caption{Selection rules for the photoluminescence of $s$-type ($m=0$) moir\'e localized IXs in (a) $R$-type and (b) $H$-type TMD heterobilayers. In all diagrams the exciton is depicted as a mid-gap level for the electron and the hole, and the emitted photon as a wavy line. Red crosses indicate that the IX state is forbidden by symmetry to decay into an in-plane-polarized photon.}
\label{fig:PLRules}
\end{center}
\end{figure*}
The six wavefunctions discussed above are plotted in Figures \ref{fig:RWSe2MoSe2wavefunction}(b)-(g). From inspection of their symmetries\cite{triangle}, we deduce that $F_{0,1}$ and $F_{0,2}$ transform according to the one-dimensional irrep $A_1$ of group $C_{3v}$. That is, $F_{0,1}$ and $F_{0,2}$ do not acquire a phase under $C_3$ rotations or mirror operations about the axes sketched in Fig.\ \ref{fig:RWSe2MoSe2wavefunction}(c). Similarly, the degenerate doublets $F_{\pm1,1}$ and $F_{\pm1,2}$ belong to the two-dimensional irrep $E$. In general, we find that degenerate states of irrep $E$ originate from the $\bar{\mm}=\pm1$ blocks, whereas the $\bar{\mm}=0$ block produces states belonging to $A_1$ and $A_2$. The first two states belonging to the latter are $F_{0,3}$ and $F_{0,6}$, which acquire a $(-1)$ phase factor under mirror operations (see Fig.\ \ref{fig:HighEnergyStates} in Appendix \ref{app:morelevels}). While IXs are---due to their permanent electric dipole moment---mainly susceptible to out-of-plane electric fields, the in-plane symmetry of the wavefunctions shown in Fig.\ \ref{fig:RWSe2MoSe2wavefunction} will determine the localized IX's interaction with perturbations such as impurities with non trivial symmetry properties, phonons, and in-plane anisotropies.

\section{Symmetry considerations and optical selection rules}\label{sec:optics} 
Having computed the RM and COM parts of the wave function, we are now in a position to construct the full excitonic state. For COM quantum numbers $(\bar{\mm},\ell)$ and RM quantum numbers $(m,n)$, and taking the bottom of the moir\'e potential well as the origin of coordinates, the localized IX wave function is given by
\begin{equation}\label{eq:IXrealspace}
\begin{split}
	\ket{{\rm IX}_{\tau',\tau}^{c',v;s}}_{\bar{\mm},\ell}^{m,n}=&\int d^2r_{\rm e}\int d^2r_{\rm h}\,e^{i(\tau'\KK'\cdot\rr_{\rm e} - \tau\KK\cdot\rr_{\rm h})}\Psi_{\bar{\mm},\ell}^{m,n}(\rr_{\rm e},\rr_{\rm h})\\
	&\times\varphi_{c',\tau',s}^\dagger(\rr_{\rm e}) \varphi_{v,\tau,s}(\rr_{\rm h})\ket{\Omega},
\end{split}
\end{equation}
where $\varphi_{\alpha,\tau,s}(\rr)$ is the field operator for an electron of band $\alpha$ with spin and valley quantum numbers $s$ and $\tau$, respectively; $\KK$ ($\KK'$) is the valley vector MX${}_2$ (M$'$X$'$${}_2$); and $| \Omega \rangle$ represents the charge-neutral many-body ground state. The two-body state \eqref{eq:IXrealspace} is formed by an electron and a hole of opposite spin projections, and thus can recombine in the absence of spin-flip mechanisms\cite{PhysRevLett.119.047401,Yu_2018}, which we shall ignore in our discussion. In addition, we shall restrict the relative values of $\tau$ and $\tau'$ in order to form exciton states that can recombine without intervalley scattering. This is achieved by setting $\tau'=\tau$ for configurations close to $R$ stacking, and $\tau'=-\tau$ for cases close to $H$ stacking. The optical activity of excitons that meet these criteria is then solely dependent on the symmetry properties of the exciton state, which we address next.

In Sec.\ \ref{sec:COM} we have identified the relation between $\bar{\mm}$ and the COM states' irreducible representation, which immediately yields the symmetry rules
\begin{equation}\label{eq:C3wf}
\begin{split}
	C_3 F_{\bar{\mm},\ell} =& e^{-i\tfrac{2\pi}{3}\bar{\mm}}F_{\bar{\mm},\ell},\\
	C_3f_{m,n}=&e^{-i\tfrac{2\pi}{3}m}f_{m,n},\\
	C_3 \Psi_{\bar{\mm},\ell}^{m,n} =& e^{-i\tfrac{2\pi}{3}(\bar{\mm}+m)}\Psi_{\bar{\mm},\ell}^{m,n},
\end{split}
\end{equation}
where the second expression stems trivially from Eq.\ \eqref{eq:fmn}. Using Eq.\ \eqref{eq:C3wf}, it can be shown that \eqref{eq:IXrealspace} transforms under $C_3$ rotations as (Appendix \ref{app:symmetry})
\begin{equation}\label{eq:C3IX}
	C_3|{\rm IX}_{\tau',\tau}^{c',v;s}\rangle_{\bar{\mm},\ell}^{m,n} = e^{-i\tfrac{2\pi}{3}(\bar{\mm}+m)}\phi_{\Omega}\phi_{c',\tau'}\phi_{v,\tau}^* |{\rm IX}_{\tau',\tau}^{c',v;s}\rangle_{\bar{\mm},\ell}^{m,n},
\end{equation}
with $\phi_{c',\tau'}$, $\phi_{v,\tau}$ and $\phi_{\Omega}$ the $C_3$ eigenvalues of the conduction- and valence-band Bloch functions and the state $| \Omega \rangle$, respectively.

The values of $\phi_{c',\tau'}$ and $\phi_{v,\tau}$ are determined by the local interlayer registry at the bottom of the potential well\cite{liu2015electronic}, encoded through the in-plane stacking vector $\rr_0$ joining the nearest metal atoms of the two layers [Figs.\ \ref{fig:MoireWSe2MoSe2}c and d]. In the case of $R$-WSe${}_2$/MoSe${}_2$ ($\tau'=\tau$), this corresponds to BA stacking ($\rr_0^{\rm BA}=-\tfrac{a_{\rm WSe_2}}{\sqrt{3}}\hat{\mathbf{y}}$), where the W atom of the WSe${}_2$ layer is aligned with the hollow site of the MoSe${}_2$ layer, while Se atoms in the WSe${}_2$ layer coincide with the Mo atoms of MoSe${}_2$. By contrast, for $H$-WSe${}_2$/MoSe${}_2$ and $H$-WSe${}_2$/MoS${}_2$  ($\tau'=-\tau$) the local stacking is BB$'$ ($\rr_0^{\rm BB'}=\tfrac{a_{\rm WSe_2}}{\sqrt{3}}\hat{\mathbf{y}}$), where once again the W atom aligns with the MoSe${}_2$ hollow site, while the chalcogens in both layers coincide. In both cases, we may take the bottom-layer W atom as the common rotation center, such that the electron Bloch function rotates about the top-layer hollow site, resulting in\cite{liu2015electronic}
\begin{subequations}\label{eq:C3Bloch}
\begin{equation}
	\phi_{v,\tau} = \exp{i\tfrac{2\pi \tau}{3}},
\end{equation}
\begin{equation}
	\phi_{c',\tau'} = \exp{-i\tfrac{2\pi\tau'}{3}}= \left\{
	\begin{array}{ccc}
	\exp{-i\tfrac{2\pi\tau}{3}} &,& R\,\text{stacking}\\
	\exp{i\tfrac{2\pi\tau}{3}} &,& H\,\text{stacking}
	\end{array}\right. .
\end{equation}
\end{subequations}
This gives for the full wave function
\begin{equation}\label{eq:C3symPsi}
	C_3|{\rm IX}_{\tau',\tau}^{c',v;s}\rangle_{\bar{\mm},\ell}^{m,n} = \left\{
	\begin{array}{ccc}
	e^{-i\tfrac{2\pi}{3}(\bar{\mm}+m-\tau)}\phi_{\Omega}|{\rm IX}_{\tau',\tau}^{c',v;s}\rangle_{\bar{\mm},\ell}^{m,n} &,& R\,\text{stacking}\\
	e^{-i\tfrac{2\pi}{3}(\bar{\mm}+m)}\phi_{\Omega}|{\rm IX}_{\tau',\tau}^{c',v;s}\rangle_{\bar{\mm},\ell}^{m,n} &,& H\,\text{stacking}
	\end{array}\right. .
\end{equation}

The optical selection rules are obtained by combining the result \eqref{eq:C3symPsi} with the symmetry properties of the many-body light-matter interaction Hamiltonian, which we derive in Appendix \ref{app:lightmatter}:
\begin{equation}\label{eq:manybodyLM}
\begin{split}
	&H_{\rm LM}=\frac{e\gamma}{\hbar c}\sum_{s,\tau}\sum_{\kk,\xxi}\sqrt{\frac{4\pi \hbar c}{SL \xi}}c_{c,\tau,s}^\dagger(\kk+\xxi_{\parallel})c_{v,\tau,s}(\kk)a_{-\tau}(\xxi)\\
	&+\frac{e\gamma'}{\hbar c}\sum_{s,\tau'}\sum_{\kk',\xxi}\sqrt{\frac{4\pi \hbar c}{SL \xi}}c_{c',\tau',s}^\dagger(\kk+\xxi_{\parallel})c_{v',\tau',s}(\kk)a_{-\tau'}(\xxi)\\
	&+\Theta(30^\circ-\theta) \frac{e\gamma_{R}}{\hbar c}\sum_{s,\tau',\tau}\,\sum_{\kk',\kk,\xxi}\sqrt{\frac{4\pi \hbar c}{SL \xi}}\vartheta_{\tau',\tau}(\kk',\kk)\\
	&\qquad\qquad\qquad\times c_{c',\tau',s}^\dagger(\kk'+\xxi_{\parallel})c_{v,\tau,s}(\kk)a_{\tau}(\xxi) + \text{H.c.}
\end{split}
\end{equation}
Here, the boson operator $a_{\eta}(\xxi)$ annihilates a single photon of wave vector $\xxi$ and in-plane circular polarization $\eta = \pm1$ (for right- and left-handed, respectively), and $\gamma$ and $\gamma'$ are proportional to matrix elements of the helical momentum operator between the conduction- and valence-band Bloch states at the valley in the corresponding layer. In the last term of \eqref{eq:manybodyLM}, $\gamma_{R}$ is the matrix elements of the same operator between bands $c'$ and $v$, and the function
\begin{equation}\label{eq:vartheta}
	\vartheta_{\tau',\tau}(\kk',\kk) = \sum_{\chi =0}^2 \delta_{\kk-\kk',C_3^\chi \Delta \KK_{\tau',\tau}} e^{i\tau C_3^\chi \KK\cdot\rr_0},
\end{equation}
containing the valley mismatch $\Delta \KK_{\tau',\tau}=\tau'\KK' - \tau\KK$ and stacking vector $\rr_0$, enforces momentum conservation\cite{anomalous_lightcones}. The Heaviside function is meant to convey that this term appears only for $R$ stacking since,  as shown in Appendix \ref{app:lightmatter}, direct optical transitions exactly at the valley are forbidden for $H$-stacked structures, and vanishingly small for crystal momenta close to the valley. When applied to moir\'e localized IXs we find that direct radiative recombination by these states is essentially absent. Therefore, photoluminescence from moir\'e localized states in $H$-type structures must take place through the indirect processes discussed in Sec.\ \ref{sec:spectrum}.

In $R$-stacked heterobilayers, direct decay of a moir\'e localized IX into a photon is determined by the matrix element
\begin{equation}
\begin{split}
	&\langle \Omega |a_{\eta}(\xxi) H_{\rm LM} |{\rm IX}_{\tau',\tau}^{c',v;s}\rangle_{\bar{\mm},\ell}^{m,n}=\\
	&\langle \Omega |a_{\eta}(\xxi)C_3^{-1}C_3 H_{\rm LM} C_3^{-1} C_3 |{\rm IX}_{\tau',\tau}^{c',v;s}\rangle_{\bar{\mm},\ell}^{m,n}=\\
	&\langle \Omega |a_{\eta}(\xxi)C_3 H_{\rm LM} C_3^{-1} |{\rm IX}_{\tau',\tau}^{c',v;s}\rangle_{\bar{\mm},\ell}^{m,n}\\
	&\qquad \times	e^{-i\tfrac{2\pi}{3}(\eta - \tau + \bar{\mm} + m)}.
\end{split}
\end{equation}
Since light-matter interactions conserve angular momentum (Appendix \ref{app:lightmatter}), $C_3 H_{\rm LM} C_3^{-1} = H_{\rm LM}$ and the matrix element can be non-zero only when the $C_3$ eigenvalue is conserved in the recombination process, leading to the optical selection rule
\begin{equation}\label{eq:SelectionRules}
	\eta = (\tau -\bar{\mm}-m)\bmod 3 \,\,, (R\,\text{stacking})
\end{equation}
illustrated in Fig.\ \ref{fig:PLRules}(a) for $m=0$. In $R$-type structures, $\bar{\mm}=0$ states at valley $\tau$ produce PL of polarization $\eta=\tau$, opposite to the well known selection rule for intralayer excitons ($\eta=-\tau$), as recently reported by Seyler \emph{et al}.\cite{seyler2019} and Tran \emph{et al}.\cite{tran2019}, whereas states with $\bar{\mm}=-\tau$ give PL of polarization $\eta=-\tau$. For $\bar{\mm}=\tau$ the IX $C_3$ eigenvalue is one, which is incompatible with the two possible values of $\eta$; these states are dark for in-plane polarized photons, and couple instead to out-of-plane polarized ones that propagate along the heterostructure plane. Since these photons are missed by most optical experiments\cite{PhysRevLett.119.047401}, we label them as ``dark''.

In the case of $H$-type structures matching the $C_3$ eigenvalues of the initial and final states gives the selection rule
\begin{equation}\label{eq:SelectionRulesH}
	\eta = -(\bar{\mm}+m)\bmod 3 \,\,, (H\,\text{stacking}),
\end{equation}
for direct recombination, where the valley quantum number is absent since the valley-dependent orbital angular momenta of the electron and hole Bloch states cancel each other out. However, as discussed above, the probability amplitude for such processes is vanishingly small. As we shall see in Sec.\ \ref{sec:spectrum}, indirect recombination mediated by interlayer tunneling obeys Eq.\ \eqref{eq:SelectionRulesH} with the additional constraint that intralayer exciton selection rules should be obeyed as well. This eventually results in different radiative or absorption rates for states with the same quantum number $\bar{\mm}$ belonging to opposite valleys. In general, $\bar{\mm}=0$ states dark, as illustrated in Fig.\ \ref{fig:PLRules}(b).


\begin{figure}[t!]
\begin{center}
\includegraphics[width=\columnwidth]{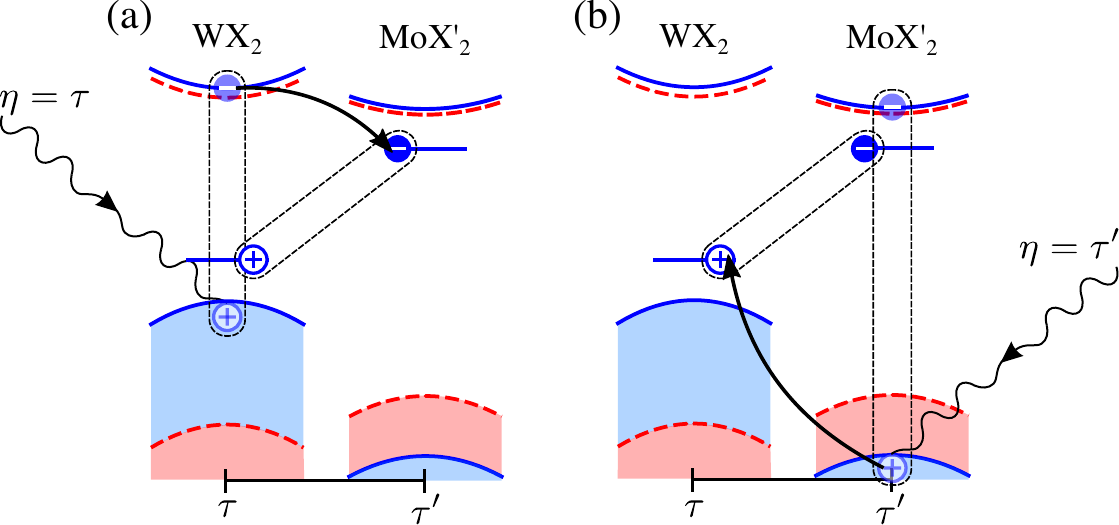}
\caption{Photon absorption processes for moir\'e localized interlayer excitons in the TMD heterobilayer $H$-WX${}_2$/MoX$'{}_2$. Bands of different spin are shown in different colors. The incoming photon (wavy line) interacts with either (a) the WX${}_2$ or (b) the MoX$'{}_2$ layer to create an intralayer exciton. In the former case, the electron may tunnel into MoX$'{}_2$ to form the localized IX state. Alternatively, in the latter case the hole may tunnel into WX${}_2$ to form the same IX state.}
\label{fig:diagramsAbs}
\end{center}
\end{figure}
\section{Absorption spectrum of moir\'e localized interlayer excitons}\label{sec:spectrum}
In Sec.\ \ref{sec:optics} we found that direct interaction of IXs with light is essentially forbidden for localized IXs in $H$-type structures. In fact even in $R$-stacked heterobilayers such interactions are weak\cite{anomalous_lightcones,topo_mosaics} due to the large spatial separation between the charge carriers forming the IX. As a result, the oscillator strength of IXs comes mainly from mixing with bright intralayer excitons\cite{anomalous_lightcones,hXprb2019} through interlayer tunneling of electrons and holes. Such mixing can be taken into account perturbatively, given the weak tunneling strengths and large interlayer CB and VB detunings\cite{gong2013band,xu2018role} typical of type-II TMD heterostructures\footnote{Notable exceptions to this rule are\cite{hXnature2019} MoSe${}_2$/WS${}_2$, and possibly\cite{hXprb2019} MoTe${}_2$/MoSe${}_2$, where strong interlayer hybridization of carriers has been predicted and/or observed experimentally. Nonetheless, these heterostructures are chalcogen mismatched and possess short moir\'e periodicities, making them poor candidates for exciton localization by moir\'e potentials.}. The dominant contributions to the photon absorption rate are given by the two processes sketched in Fig.\ \ref{fig:diagramsAbs}, where an incoming photon, depicted as a wavy line, is shown to interact with either of the two TMD layers to create a \emph{virtual} intralayer exciton. Then, one of the two carriers can tunnel into the opposite layer to form a moir\'e localized IX, shown as mid-gap levels for the electron and the hole.

This process introduces additional constraints on top of the selection rules obtained in Sec.\ \ref{sec:optics}. Firstly, formation of the intermediate exciton state in either layer is constrained by the intralayer optical selection rule $\eta=-\tau$ if it forms in the MX${}_2$ layer, or $\eta=-\tau'$ if it forms in the M$'$X$'{}_2$ layer. Moreover, the intermediate exciton must have an angular momentum quantum number\cite{chernikov2014} $m=0$, which sets the same quantum number for the moir\'e localized IX. In other words, $s$-type IXs will dominate the absorption spectrum associated to moir\'e localized excitons.

To compute the probability amplitudes of the processes in Fig.\ \ref{fig:diagramsAbs}, we first evaluate the corresponding mixed intralayer-interlayer exciton wavefunction, to first order in perturbation theory:
\begin{equation}\label{eq:hX}
\begin{split}
	|& \mathrm{hX}_{\tau',\tau}^{c',v;s} \rangle_{\bar{\mm},\ell}^{0,n}=| \mathrm{IX}_{\tau',\tau}^{c',v;s} \rangle_{\bar{\mm},\ell}^{0,n}\\
	&+ \sum_{\QQ,\bar{n}}\frac{{}_{\bar{n}}\langle \mathrm{X}_{\tau,\tau}^{c,v;s}(\QQ) | H_{T} | \mathrm{IX}_{\tau',\tau}^{c',v;s}\rangle_{\bar{\mm},\ell}^{0,n}}{E_{\bar{\mm},\ell,\tau}^{0,n;s}-E_{\tau,\bar{n};s}(Q)}| \mathrm{X}_{\tau,\tau}^{c,v;s}(\QQ) \rangle_{\bar{n}}\\
	&+ \sum_{\QQ,\bar{n}}\frac{{}_{\bar{n}}\langle \mathrm{X}_{\tau',\tau'}^{c',v';s}(\QQ) | H_T | \mathrm{IX}_{\tau',\tau}^{c',v;s}\rangle_{\bar{\mm},\ell}^{0,n}}{E_{\bar{\mm},\ell,\tau}^{0,n;s}-E_{\tau',\bar{n};s}'(Q)}| \mathrm{X}_{\tau',\tau'}^{c',v';s},(\QQ) \rangle_{\bar{n}},
\end{split}
\end{equation}
where $|{\rm X}_{\tau,\tau}^{c,v}(\QQ) \rangle_{\bar{n}}$ and $|{\rm X}_{\tau',\tau'}^{c',v'}(\QQ) \rangle_{\bar{n}}$ are the wave functions of $s$-type intralayer excitons with COM wave vector $\QQ$ in MX${}_2$ and M$'$X$'{}_2$, respectively. We have also introduced the intralayer exciton dispersions
\begin{equation}\label{eq:Xdispersions}
\begin{split}
	E_{\tau,\bar{n};s}(Q) =& E_{\tau s;\bar{n}}^0 + \frac{\hbar^2 Q^2}{2(m_{\rm e}+m_{\rm h})},\\
	E_{\tau',\bar{n};s}'(Q) =& E_{\tau's;\bar{n}}^{'0} + \frac{\hbar^2 Q^2}{2(m_{\rm e'}+m_{\rm h'})},
\end{split}
\end{equation}
with $m_{\rm e}$ and $m_{\rm h'}$ are the electron and hole masses in the MX${}_2$ and M$'$X${}_2'$ layers, respectively.
The interlayer tunneling Hamiltonian for the moir\'e heterostructure is\cite{hongyi2017_2, hXprb2019}
\begin{equation}\label{eq:Htunnel}
\begin{split}
	H_T = \sum_{s,\tau,\tau'}&\sum_{\kk,\kk'}\vartheta_{\tau',\tau}(\kk',\kk)\Big[t_c c_{c',\tau',s}^\dagger(\kk')c_{c,\tau,s}(\kk)\\
	 &+ t_v \exp{i\tfrac{2\pi\eta}{3}(\tau'-\tau)}c_{v',\tau',s}^\dagger(\kk') c_{v,\tau,s}(\kk)\Big] + \text{H.c.},
\end{split}
\end{equation}
where $t_{c}$ and $t_{v}$ are hopping parameters.
\begin{table}[t!]
\caption{Energies of $A$ ($\tau s = -1$) and $B$ ($\tau s = +1$) intra- and interlayer excitons in WSe${}_2$/MoSe${}_2$, relevant to photon absorption through the processes of Fig.\ \ref{fig:diagramsAbs}. The $A$-exciton energies were extracted from Refs.\ \onlinecite{seyler2019} and \onlinecite{tran2019}. The $B$-exciton values were estimated by adding the experimental spin-orbit splittings reported in Ref.\ \onlinecite{ArpesMasses} to the $A$-exciton energies.}
\begin{center}
\begin{tabular}{c r@{.}l r@{.}l r@{.}l | r@{.}l r@{.}l r@{.}l}
\hline\hline
 & \multicolumn{6}{c}{$A$ exc.\ energy [eV]} & \multicolumn{6}{c}{$B$ exc.\ energy [eV]}\\
 & \multicolumn{2}{c}{$\bar{n}=1$} & \multicolumn{2}{c}{$\bar{n}=2$} & \multicolumn{2}{c}{$\bar{n}=3$}  & \multicolumn{2}{c}{$\bar{n}=1$} & \multicolumn{2}{c}{$\bar{n}=2$} & \multicolumn{2}{c}{$\bar{n}=3$}\\
 \hline
WSe${}_2$     & 1&715${}^{\rm a,b}$ & 1&837 & 1&872 &     2&163${}^{\rm a,b}$ & 2&285 & 2&320\\
MoSe${}_2$   & 1&630${}^{\rm a}$    & 1&758 & 1&796 &     1&871 & 1&999 & 2&037\\
\hline\hline
 & \multicolumn{6}{c}{$R$ stacking [eV]}  & \multicolumn{6}{c}{$H$ stacking [eV]}\\
 \hline
IX                   &  \multicolumn{6}{c}{1.320${}^{\rm a}$} & \multicolumn{6}{c}{1.392${}^{\rm a}$}\\
\hline\hline
\end{tabular}

${}^{\rm a}$Reference [\onlinecite{seyler2019}], ${}^{\rm b}$Reference [\onlinecite{tran2019}]
\end{center}
\label{tab:Xvalues}
\end{table}%

$E_{\tau s; \bar{n}}^{0}$ and $E_{\tau s;\bar{n}}^{'0}$ in Eq.\ \eqref{eq:Xdispersions} are given by the corresponding intralayer band gaps and binding energies, as obtained in Sec.\ \ref{sec:RM}. Instead, we choose to extract the values of $E_{\tau s; 1}^0$ and $E_{\tau s; 1}^{'0}$ from experiments\cite{seyler2019,tran2019}: for $\tau s = -1$, these correspond to the $A$ exciton in the respective layer, whereas for $\tau s = 1$ the energies refer to the $B$ exciton. The full series of states can then be estimated by combining these experimental values with the level spacings predicted by our method of Sec.\ \ref{sec:RM}. We follow the same strategy for the IX, writing $E_{\bar{\mm},\ell,\tau}^{0,n;s} = E_{\tau',\tau;s}^0 + E_F^{\bar{\mm},\ell}$. The reference energy $E_{\tau',\tau;s}^0$ was taken from Ref.\ \onlinecite{seyler2019}, using the lowest measured localized state energy for $R$ stacking as $E_{1,1;\downarrow}$, and the same for $H$ stacking as $E_{-1,1;\downarrow}$. The compiled results are shown in Table \ref{tab:Xvalues}, including intralayer exciton states with $\bar{n}=1,2$ and $3$.

Next, we compute the decay rate of a single-photon state $|\eta,\xxi \rangle = a_{\eta}^\dagger(\xxi)|\Omega \rangle$ into every possible weakly hybridized exciton \eqref{eq:hX} with Fermi's golden rule
\begin{equation}\label{eq:decay}
	\Gamma_\eta(\xxi) = \frac{2\zeta}{\hbar}\sum_{\bar{\mm},\ell,n}\sum_{\tau,s}\frac{\left| \langle \eta, \xxi | H_{\rm LM} | {\rm hX}_{\tau',\tau}^{c',v;s} \rangle_{\bar{\mm},\ell}^{0,n} \right|^2}{\left(\hbar c\xi - E_{\bar{\mm},\ell,\tau}^{0,n;s}\right)^2 + \zeta^2},
\end{equation}
with a relaxed energy conservation condition to allow for a phenomenological Lorentzian broadening $\zeta$ caused by impurities and disorder. 
The absorption rate $A_{\eta}(\varepsilon)$ for photons of energy $\varepsilon$ and circular polarization $\eta$ can be estimated by multiplying \eqref{eq:decay} by the number of localizing centers per unit area---one per moir\'e unit cell---and by the number of available photon states within a range $\Delta\varepsilon$ of that energy\cite{hXprb2019}, which may be identified with the energy resolution of the measurement. For simplicity, we consider only $1s$ states for both the localized IX and the virtual intralayer excitons [$\bar{n}=n=1$ in the sum of Eq.\ \eqref{eq:decay}]. This gives
\begin{widetext}
\begin{equation}\label{eq:Arate}
	A_{\eta}(\varepsilon) = \frac{e^2}{\hbar c} \frac{16\varepsilon\,\Delta\varepsilon}{\sqrt{3}a_{\rm M}^2\hbar^3c^2}  \sum_{s,\tau}\sum_{\bar{\mm},\ell} \Bigg| \sum_{\mu=0}^2\Phi_{\bar{\mm},\ell}^{\tau',\tau,\mu}\Bigg(\gamma\, t_c \frac{X_{\tau,\tau;1}^{c,v\,*}(0) I_{\tau}}{E_{\bar{\mm},\ell,\tau}^{0,1;s} - E_{\tau s;1}^0}\delta_{\eta,-\tau} - \gamma' t_v \frac{e^{i\tfrac{2\pi}{3}(\tau-\tau')\mu}X_{\tau',\tau';1}^{c',v'\,*}(0) I_{\tau'}'}{E_{\bar{\mm},\ell,\tau}^{0,1;s} - E_{\tau' s;1}^{'0}}\delta_{\eta,-\tau'} \Bigg) \Bigg|^2 \frac{1}{\pi}\frac{\zeta}{(\varepsilon - E_{\bar{\mm},\ell,\tau}^{0,1;s})^2+\zeta^2},
\end{equation}
\end{widetext}
where $X_{\tau,\tau;\bar{n}}^{c,v}(\rrho)$ and $X_{\tau',\tau';\bar{n}}^{c',v'}(\rrho)$ are the MX${}_2$ and M$'$X$'{}_2$ intralayer exciton real-space RM wave functions, and we have defined
\begin{equation}
\begin{split}
	\Phi_{\bar{\mm},\ell}^{\tau',\tau,\mu} =& \tilde{F}_{\bar{\mm},\ell}(-C_3^\mu \Delta \KK_{\tau',\tau})e^{-i\tau C_3^\mu \KK\cdot\rr_0},\\
	I_{\tau} =& \int d^2\rho\,X_{\tau,\tau;1}^{c,v\,*}(\rrho)f_{0,1}(\rrho),\\
	I_{\tau'}' =& \int d^2\rho\,X_{\tau',\tau';1}^{c',v'\,*}(\rrho)f_{0,1}(\rrho).
\end{split}
\end{equation}
The intra- and interlayer exciton RM wave functions were computed using the numerical method of Sec.\ \ref{sec:RM}, and numerically integrated to evaluate $I_{\tau}$ and $I_{\tau'}'$ .

\subsection{Optical selection rules for absorption}\label{sec:ABSRrules}
Equation \eqref{eq:Arate} encodes the optical selection rules for absorption through the processes of Fig.\ \ref{fig:diagramsAbs}. We begin with the case of $H$-stacked structures, where the individual layer contributions to absorption are
\begin{subequations}
\begin{equation}\label{eq:APabsorptionX}
\begin{split}
	A_{\eta=-\tau}^{(H)}=&\frac{e^2}{\hbar c}\frac{16\varepsilon \Delta \varepsilon |\gamma|^2t_c^2|X_{\tau,\tau;1}^{c,v *}(0)|^2}{\sqrt{3} a_{\rm M}^2h^3c^2}\sum_{\bar{\mm},\ell,s}\left|\frac{I_{\tau}}{E_{\bar{\mm},\ell,\tau}^{0,1;s}-E_{\tau s; 1}^0}\right|^2\\
	&\times \left|\sum_{\mu=0}^2\Phi_{\bar{\mm},\ell}^{-\tau,\tau,\mu}\right|^2\frac{\zeta/\pi}{(\varepsilon - E_{\bar{\mm},\ell,\tau}^{0,1;s})^2+\zeta^2},
\end{split}
\end{equation}
\begin{equation}\label{eq:APabsorptionXp}
\begin{split}
	A_{\eta=\tau}^{(H)}=&\frac{e^2}{\hbar c}\frac{16\varepsilon \Delta \varepsilon |\gamma'|^2t_v^2|X_{\tau,\tau;1}^{c',v' *}(0)|^2}{\sqrt{3} a_{\rm M}^2h^3c^2}\sum_{\bar{\mm},\ell}\left|\frac{I_{-\tau}'}{E_{\bar{\mm},\ell,\tau}^{0,1;s}-E_{-\tau s;1}^{'0}}\right|^2\\
	&\times\left|\sum_{\mu=0}^2\Phi_{\bar{\mm},\ell}^{-\tau,\tau,\mu}e^{-i\tfrac{2\pi}{3}\tau\mu}\right|^2\frac{\zeta/\pi}{(\varepsilon - E_{\bar{\mm},\ell,\tau}^{0,1;s})^2+\zeta^2}.
\end{split}
\end{equation}
\end{subequations}
In each case, the sum over $\mu$ can be simplified by noting that $\tilde{F}_{\bar{\mm},\ell}(C_3\qq)=e^{i\tfrac{2\pi}{3}\bar{\mm}}\tilde{F}_{\bar{\mm},\ell}(\qq)$, yielding the same rule for $\Phi_{\bar{\mm},\ell}^{\tau',\tau}$. This gives
\begin{subequations}
\begin{equation}\label{eq:APselectX}
\begin{split}
	\sum_{\mu=0}^2 \Phi_{\bar{\mm},\ell}^{-\tau,\tau,\mu} =& \tilde{F}_{\bar{\mm},\ell}(-\Delta \KK_{-\tau,\tau})\\
	&\times\left[1 + 2\cos{\left\{\tfrac{2\pi}{3}(\bar{\mm}-\tau) \right\}} \right],
\end{split}
\end{equation}
\begin{equation}\label{eq:APselectXp}
\begin{split}
	\sum_{\mu=0}^2 \Phi_{\bar{\mm},\ell}^{-\tau,\tau,\mu}e^{-i\tfrac{2\pi}{3}\tau\mu} =& \tilde{F}_{\bar{\mm},\ell}(-\Delta \KK_{-\tau,\tau})\\
	&\times\left[1 + 2\cos{\left\{\tfrac{2\pi}{3}(\bar{\mm}+\tau) \right\}} \right].
\end{split}
\end{equation}
\end{subequations}
Expressions \eqref{eq:APselectX} and \eqref{eq:APselectXp} give a finite contribution to absorption by the MX${}_2$ layer only for $\bar{\mm}=\tau=-\eta$, whereas the M$'$X$'{}_2$ layer contributes only for $\bar{\mm}=\tau'=-\eta$, and $\bar{\mm}=0$ states are dark, in full agreement with Eq.\ \eqref{eq:SelectionRulesH}. Note that the absorption rates \eqref{eq:APabsorptionX} and \eqref{eq:APabsorptionXp} are quantitatively different, given the different intralayer exciton energies and wave functions at the two layers, as well as the different tunneling energies $t_{\rm c}$ and $t_{\rm v}$. Since each of these rates is associated with a specific valley polarization, their difference translates into a valley imbalance, even though the valley quantum numbers did not appear in the basic selection rule \eqref{eq:SelectionRulesH}.

\begin{figure}[h!]
\begin{center}
\includegraphics[width=0.8\columnwidth]{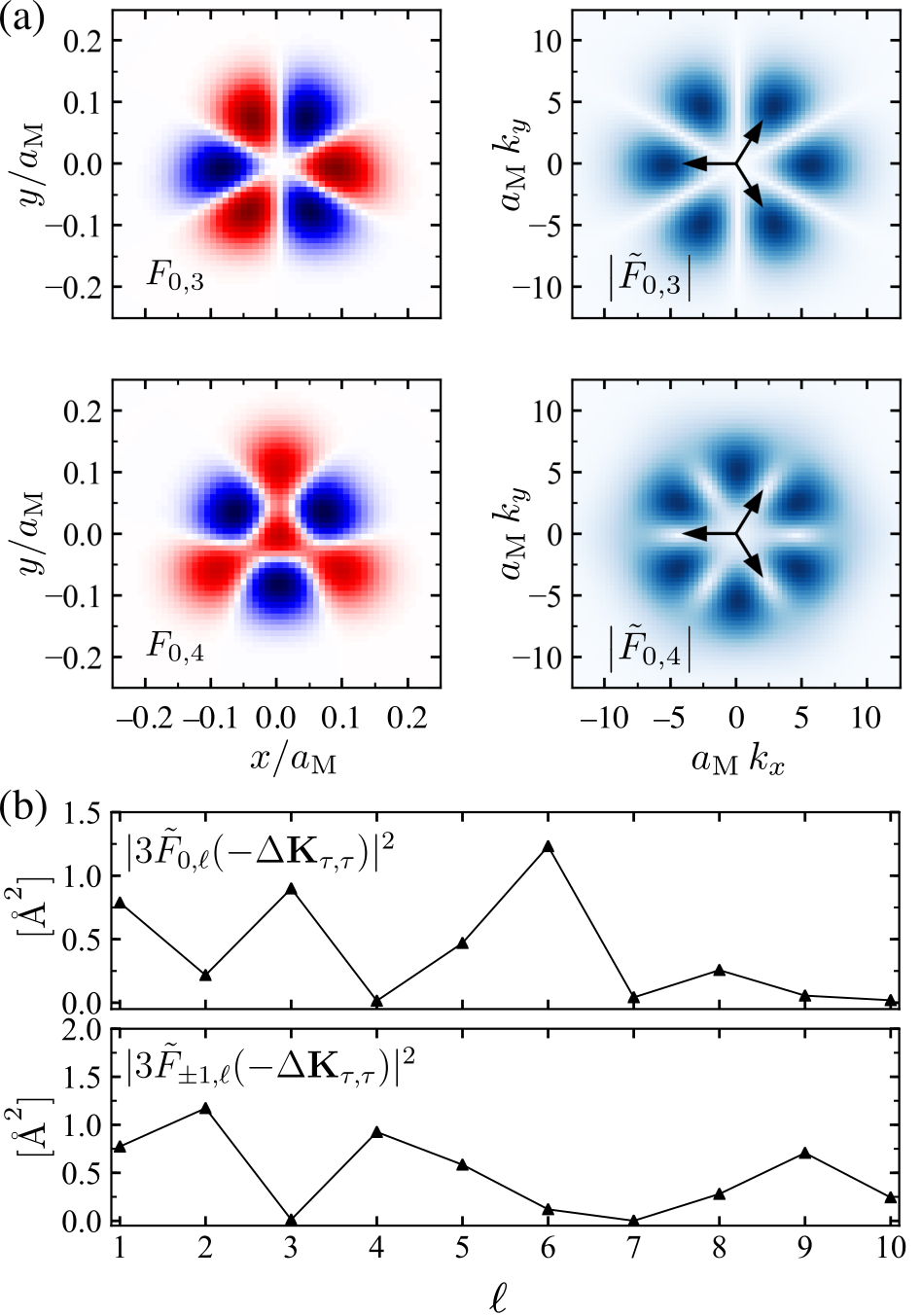}
\caption{(a) Moir\'e localized IX wavefunctions $F_{0,\ell}$ and the moduli of their Fourier transforms $|\tilde{F}_{0,\ell}|$ for $\ell=2$ and $3$ in $R$-WSe${}_2$/MoSe${}_2$. The arrows plotted in the right panels correspond to the three vectors $-C_3^\mu \Delta \KK_{\tau,\tau}$, with $\mu=0,1,2$ and $\tau=1$. (b) The weighting factor $|\tilde{F}_{\bar{\mm},\ell}(-\Delta\KK_{1,1})|^2$ as a function of $\ell$ for $\bar{\mm}=0$ (top) and $\mm=1$ (bottom), evaluated for the same heterobilayer.}
\label{fig:FTH0}
\end{center}
\end{figure}

For $R$ stacking the two layer-polarized processes interfere to give the absorption rate
\begin{equation}
\begin{split}
	A_{\eta=-\tau}^{(R)}(\varepsilon) =& \frac{e^2}{\hbar c}\frac{16\varepsilon \Delta\varepsilon}{\sqrt{3} a_{\rm M} h^3c^2}\sum_{\bar{\mm},\ell,s}\left| \sum_{\mu=0}^2\Phi_{\bar{\mm},\ell}^{\tau,\tau,\mu} \right|^2\frac{\zeta/\pi}{(\varepsilon - E_{\bar{\mm},\ell,\tau}^{0,1;s})^2+\zeta^2}\\
	&\times \left| \gamma\, t_c \frac{X_{\tau,\tau;1}^{c,v\,*}(0) I_{\tau}}{E_{\bar{\mm},\ell,\tau}^{0,1;s} - E_{\tau,1;s}^0} - \gamma' t_v \frac{X_{\tau,\tau;1}^{c',v'\,*}(0) I_{\tau}'}{E_{\bar{\mm},\ell,\tau}^{0,1;s} - E_{\tau,1;s}'{}^0} \right|^2,
\end{split}
\end{equation}
where we have
\begin{equation}\label{eq:Pselect}
\begin{split}
	\sum_{\mu=0}^2\Phi_{\bar{\mm},\ell}^{\tau,\tau,\mu} = \tilde{F}_{\bar{\mm},\ell}(-\Delta\KK_{\tau,\tau})\left[1+2\cos{\left\{\tfrac{2\pi}{3}(\bar{\mm}+\tau) \right\}} \right].
\end{split}
\end{equation}
Expression \eqref{eq:Pselect} is non zero only for $\bar{\mm}=-\tau=\eta$ in agreement with Eq.\ \eqref{eq:SelectionRules}, corresponding to the inverse process to that in the bottom panel of Fig.\ \ref{fig:PLRules}(a). Though allowed by symmetry in $R$-stacked structures, absorption by $\bar{\mm}=0$ states cannot be mediated by intralayer excitons, as it violates the intralayer optical selection rules.

\begin{figure}[ht!]
\begin{center}
\includegraphics[width=0.8\columnwidth]{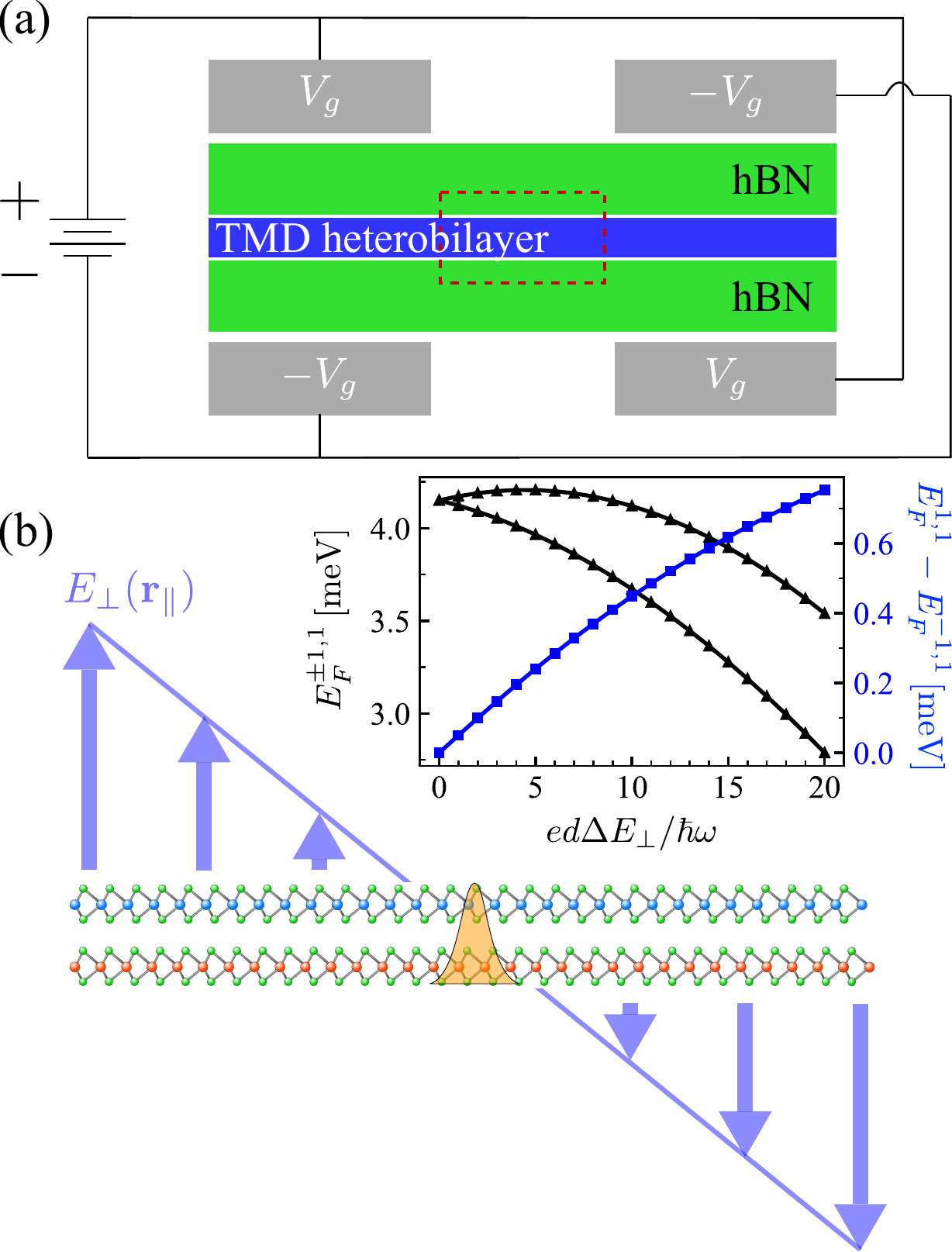}
\caption{(a) Proposed experimental setup to test symmetry properties of $\bar{\mm}=\pm1$ moir\'e localized IXs. (b) An out-of-plane electric field $E_\perp$ whose intensity varies across the moir\'e supercell is produced in the central region of the split double gate [dashed rectangle in panel (a)]. The inset shows the modulation of the $(\ell=1$, $\bar{\mm}=\pm1)$ states (left axis) and their detuning (right axis), as a function of the in-plane gradient of the out-of-plane electric field.}
\label{fig:expfig}
\end{center}
\end{figure}

The above analysis leads to the important conclusion that the dominant absorption processes favor localized IXs of the $E$ irrep of group $C_{3v}$. Previous works\cite{anomalous_lightcones,hongyi2017,Yu_2018} have shed light on the optical selection rules of extended IXs in TMD heterobilayers. In particular, Ref.\ \onlinecite{Yu_2018} gives a thorough account of the spatially-dependent selection rules for the interaction between in-plane isotropic IX wave packets and in- and out-of plane polarized light within the moir\'e supercell, based on the symmetry of the electron and hole Bloch states. Nonetheless, our theory shows that the orbital degrees of freedom of the localized IX state play a crucial role in its interaction with light. Indeed, the selection rules of Eqs.\ \eqref{eq:SelectionRules} and \eqref{eq:SelectionRulesH} for $(\bar{\mm}=0,\,\ell=1)$ states coincide with those reported in Refs.\ \onlinecite{hongyi2017} and \onlinecite{Yu_2018} since those states transform as scalars under $C_{3v}$ operations, just like an isotropic wave packet would.

Our results suggest a strong coupling to light for states with $\bar{\mm}=\pm1$, whose symmetry properties make them susceptible to perturbations that single our any particular direction along the sample plane. To see this, note that such a perturbation $H_{\rm P}$ transforms like a vector under $C_{3v}$ operations, and thus belongs to irrep $E$.  Straightforward group theory arguments\cite{dresselhaus2007group} show that $H_{\rm P}$ then couples the states $\bar{\mm}=1$ and $\bar{\mm}=-1$, lifting the degeneracy of all doublets in the localized IX spectrum, and in general modifying their optical spectrum.

To be specific, let us consider an out-of-plane electric field $E_\perp$ whose magnitude varies linearly along some arbitrary direction $\hat{\mathbf{v}}$ on the sample plane. This field couples to the permanent electric dipole of IXs, producing a spatially dependent Stark shift
\begin{equation}\label{eq:stark}
	H_{\rm P}=\frac{ed\Delta E_\perp}{a_M}\RR\cdot \hat{\mathbf{v}},
\end{equation}
where $\Delta E_\perp$ gives the change in the electric field magnitude over a distance equal to the moir\'e superlattice constant along vector $\hat{\mathbf{v}}$, and $\RR$ is the COM position. Experimentally, this may be achieved in the split double-gate setup of Fig.\ \ref{fig:expfig}(a), where opposite voltage drops are applied to the left and right segments to produce an electric field that varies in the central region along a single direction. The inset of Fig.\ \ref{fig:expfig}(b) shows the splitting of the lowest-energy doublet $E_{F}^{\pm1,1}$ in the moir\'e localized IX spectrum of aligned $R$-WSe${}_2$/MoSe${}_2$ produced by the perturbation \eqref{eq:stark}, as computed by direct diagonalization. Our calculations predict a splitting of the order of the level spacing ($\hbar \omega = 2.2\,{\rm meV}$ in this case) for a modest electric field gradient of $5.7\times10^{-3} {\rm mV}\cdot\text{\AA}^{-2}$; that is, an electric field change of $10\,{\rm mV}/\text{\AA}$ over a moir\'e supercell. This effect could be used to experimentally verify our predictions for the spectrum and selection rules of moir\'e localized IXs.


\subsection{Numerical results}
\begin{figure*}[t!]
\begin{center}
\includegraphics[width=1.8\columnwidth]{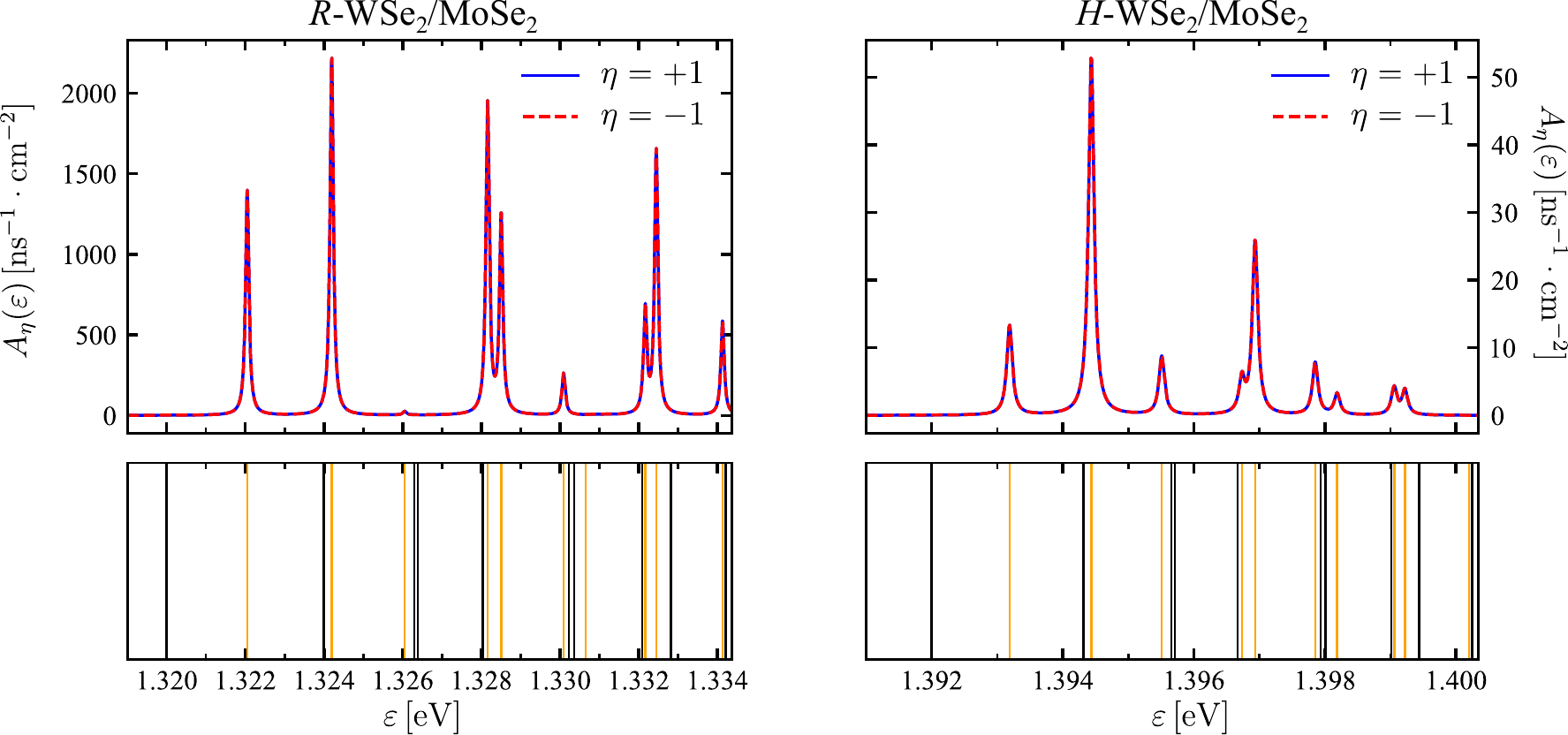}
\caption{Absorption rate per unit area of moir\'e localized IXs in $R$-WSe${}_2$/MoSe${}_2$ (left panel) and $H$-WSe${}_2$/MoSe${}_2$ (right panel), for right- ($\eta=+1$) and left-handed ($\eta=-1$) polarized light.  The bottom panels show the corresponding calculated IX spectra, with $\bar{\mm}=0$ ($\bar{\mm}=\pm1$) states represented by black (orange) lines. The absolute IX energies are based on the experimental values reported in Table \ref{tab:Xvalues} combined with our numerical results, and we have used phenomenological $\beta=50\,\mu{\rm eV}$ line widths for the absorption resonances, a detector resolution $\Delta\varepsilon=50\,\mu{\rm eV}$, and hopping strengths $t_{\rm v}=2t_{\rm c}=52\,{\rm meV}$.}
\label{fig:AbsRH}
\end{center}
\end{figure*}

Equations \eqref{eq:APselectX}, \eqref{eq:APselectXp} and \eqref{eq:Pselect} show that the localized IX's oscillator strength is determined by the Fourier transform of the COM wave function evaluated at a single wave vector $-\Delta\KK_{\tau',\tau}$. Note that, while the latter vector is determined by the moir\'e supercell vectors [Fig.\ \ref{fig:MoireWSe2MoSe2}(a)], the wave function $F_{\bar{\mm},\ell}$ is given by the trigonally symmetric potential well, whose orientation in the moir\'e pattern is given by the phase factor $\varphi$ appearing in Eq.\ \eqref{eq:fit}.

Figure \ref{fig:FTH0}(a) shows two sample functions $\tilde{F}_{0,3}$ and $\tilde{F}_{0,4}$ for $R$-WSe${}_2$/MoSe${}_2$, with arrows indicating the vectors $-C_3^\mu \Delta \KK_{1,1}$. In the former case the three vectors coincide with lobes of the function, giving a finite oscillator strength, whereas in the latter they align with nodes, resulting in a dark state. Figure \ref{fig:FTH0}(b) shows the corresponding prefactors $|3\tilde{F}_{\bar{\mm},\ell}(-\Delta\KK_{1,1})|^2$ for the first few localized IXs, indicating that the states with $\bar{\mm}=0$ and $\ell=7$ and $\ell=10$, and those with $\bar{\mm}=\pm1$ and $\ell=3,\,7$ and $10$ are also dark. This effect is, in a sense, accidental, as it is independent of the potential well symmetry. Further \emph{ab initio} studies of TMD heterostructures are necessary to ascertain whether different pairs of TMDs will produce the same moir\'e potential landscape for IXs when similarly stacked into a heterostructure. Here, we simply focus on the cases of $R$- and $H$-stacked WSe${}_2$/MoSe${}_2$, which we consider to be of interest for experiments.

Figure \ref{fig:AbsRH} shows the polarization-resolved absorption spectra of moir\'e localized IXs in $R$- and $H$-stacked WSe${}_2$/MoSe${}_2$. The curves were constructed based on the selection rules of Sec.\ \ref{sec:ABSRrules}, our numerical results for $\tilde{F}_{\bar{\mm},\ell}(-\Delta\KK_{\tau',\tau})$, and the experimental data of Table \ref{tab:Xvalues}. For the tunnelling strengths entering Eq.\ \eqref{eq:Htunnel}, we use $t_{\rm v} = 2t_{\rm c} = 52\,{\rm meV}$ motivated by earlier work on TMD heterobilayers\cite{hXnature2019,hXprb2019}, which should suffice for an order-of-magnitude estimation. The corresponding localized IX spectra are shown in the bottom panels to highlight the absence of absorption signatures from $\bar{\mm}=0$ states, as dictated by symmetry, as well as from some $\bar{\mm}\ne0$ states for which $F_{\bar{\mm},\ell}(-\Delta \KK_{\tau',\tau})\approx0$. The absorption profiles for right- and left-circularly polarized light are identical, as guaranteed by time reversal symmetry.

\section{Conclusions}\label{sec:conclusions}
We have presented an in-depth study of interlayer exciton localization by trigonally symmetric moir\'e potentials in type-II transition-metal dichalcogenide heterostructures. Using to our advantage the natural scale separation between the exciton's center of mass and relative motions in heterobilayers with large moir\'e supercells, we have applied direct diagonalization techniques to predict the localized exciton spectra, and to identify each state's symmetry properties as inherited from the localizing potential and the carrier Bloch functions. We have demonstrated that the \emph{ab initio} potentials proposed by Yu \emph{et al.}\cite{hongyi2017} produce exciton spectra which are consistent with the experimental results of Seyler \emph{et al.}\cite{seyler2019} Moreover, our techniques can be easily adapted to more realistic potentials, including lattice relaxation effects\cite{enaldievPRL,westonNatNano,rosenberger2020}.

We have derived general optical selection rules for moir\'e localized exciton states which explicitly include their orbital center of mass motion, thus generalizing earlier predictions\cite{hongyi2017,Yu_2018} based solely on the symmetries of carrier Bloch functions. By pairing these selection rules with the dominant exciton-photon interaction processes, we have predicted that the moir\'e-localized-state sector of the optical absorption in these materials is dominated by degenerate pairs of states belonging to the $E$ irreducible representation of the group $C_{3v}$. Due to their symmetry, the degeneracy of these states can be lifted by any perturbation that singles out a specific in-plane direction in the heterostructure. We have proposed such a perturbation in the form of an out-of-plane electric field whose magnitude is modulated along an arbitrary direction within the plane, which we believe can be accomplished in a split gate setup (Fig.\ \ref{fig:expfig}). Based on our numerical results for localized interlayer excitons in WSe${}_2$/MoSe${}_2$, we expect that a modest electric field variation of only $10\,{\rm mV}\cdot\text{\AA}^{-1}$ over a moir\'e supercell should split the first degenerate doublet in the spectrum by $\sim 1\,{\rm meV}$, allowing to verify our predictions for the symmetry and optical selection rules of these states. The results presented in this paper are fundamental to understanding how general perturbations influence moir\'e localized exciton states, and lay the groundwork for the manipulation of their optical response.

\acknowledgments{All authors acknowledge funding from DGAPA-UNAM, through projects PAPIIT no.\ IN113920 and PAPIME PE107219.  D.A.R-T.\  and F.M. acknowledge the support of DGAPA-UNAM through its postdoctoral fellowships program. D.A.R-T.\ thanks M.\ Danovich and D.\ Perello for many fruitful discussions on the present subject. I.S. thanks the hospitality of CNyN-UNAM during the event \emph{Escuela Nacional de Nanociencias 2019}, where this project was conceived.}

\appendix

\section{Effective interactions between carriers in a semiconducting bilayer}\label{app:interaction}
The semiclassical electrostatic interaction between carriers in a semiconducting heterobilayer is derived by considering charges moving in two infinitely thin dielectric films of susceptibilities $\kappa_1$ and $\kappa_{-1}$ separated along the $z$ direction, embedded in an environment with average dielectric constant $\epsilon$. The analysis below is reproduced from Ref.\ \onlinecite{danovich_2018}.

Assume that layer $j$, located at $z=z_j$, contains a charge distribution $\rho^j(\rr,z) = \rho^j(\rr)\delta(z-z_j)$, where $\rr$ is an in-plane vector. Gauss's law relates this charge density to the electrostatic potential $\phi(\rr,z)$ as\cite{landauEM}
\begin{equation}\label{eq:gauss}
\begin{split}
	\rho^j(\rr)\delta(z-z_j) =& -\frac{\epsilon}{4\pi}\nabla^2\phi(\rr,z)\\
	&-\sum_{\ell=1,-1}\kappa_\ell(\partial_x^2+\partial_y^2)\phi(\rr,z_\ell)\delta(z-z_\ell).
\end{split}
\end{equation}
Taking the three-dimensional Fourier transform of \eqref{eq:gauss}, followed by the inverse Fourier transform only in the $x$ and $y$ directions, we obtain
\begin{equation}\label{eq:gauss_fourier}
	\rho^j(\qq)e^{-q|z-z_j|}=\frac{q\epsilon}{2\pi}\phi(\qq,z)+q^2\sum_{\ell=1,-1}\kappa_\ell\phi(\qq,z_\ell)e^{-q|z-z_\ell|},
\end{equation}
where $\qq$ is an in-plane wave vector and
\begin{equation}
	g(\qq,z)=\int d^2r \,e^{-i\qq\cdot\rr}g(\rr,z).
\end{equation}
Equation \eqref{eq:gauss_fourier} can be evaluated within a given semiconducting layer $n$ by setting $z=z_n$, leading to the coupled equations
\begin{subequations}
\begin{equation}
	\rho^j(\qq) = q\left(\frac{\epsilon}{2\pi}-\kappa_jq \right)\phi(\qq,z_j),\quad (n=j),
\end{equation}
\begin{equation}
\begin{split}
	\rho^j(\qq)e^{-qd} =& q\left(\frac{\epsilon}{2\pi}-\kappa_{-j}q \right)\phi(\qq,z_j)\\
	&+q^2\kappa_{j}\phi(\qq,d_j)e^{-qd},\quad (n=-j),
\end{split}
\end{equation}
\end{subequations}
where we have introduced the vertical distance between the films, $d$. Solving the system for the potential within layer $j$ gives $\phi(\qq,z_j) = U_j^{\rm intra}(\qq)\rho^j(\qq)$, with the intralayer interaction
\begin{equation}\label{eq:Vintra}
	U_j^{\rm intra}(q) = \frac{2\pi [1+r_*^{(-j)}q(1-e^{-2dq})]}{\epsilon q [1+q(r_*^{(j)}+r_*^{(-j)}) + q^2 r_*^{(j)}r_*^{(-j)}(1-e^{-2dq}) ]},
\end{equation}
and the definition $r_*^{(j)}=\tfrac{2\pi \kappa_j}{\epsilon}$. Conversely, solving the system of equations for the potential in the opposite layer to that containing the charge distribution we find $\phi(\qq,z_{-j}) = U_j^{\rm inter}(\qq)\rho^j(\qq)$, with the interlayer interaction
\begin{equation}\label{eq:Vinter}
	U_j^{\rm inter}(q) = \frac{2\pi }{\epsilon q[(1+qr_*^{(j)})(1+qr_*^{(-j)})e^{dq} - q^2r_*^{(j)}r_*^{(-j)}e^{-dq}]}.
\end{equation}
Neither \eqref{eq:Vintra} nor \eqref{eq:Vinter} have analytical expressions in real space, since their inverse Fourier transforms cannot be computed in closed form. However, in the long-range approximation ($qr_*^{(j)}\ll1$ for either $j$) these expressions become
\begin{subequations}
\begin{equation}\label{eq:VintraLR}
	\left. U_j^{\rm intra}(q) \right|_{qr_*^{(j)}\ll1} = \frac{2\pi }{\epsilon q [1+q(r_*^{(j)}+r_*^{(-j)})]},
\end{equation}
\begin{equation}\label{eq:VinterLR}
	\left. U_j^{\rm inter}(q) \right|_{qr_*^{(j)}\ll1} = \frac{2\pi }{\epsilon q[1+q(r_*^{(j)}+r_*^{(-j)} + d)]},
\end{equation}
\end{subequations}
whose inverse Fourier transforms are
\begin{subequations}
\begin{equation}\label{eq:VintraLRreal}
	U_j^{\rm intra}(r) = \frac{\pi}{2\epsilon r_{\rm eff}^{\rm intra}}\left[H_0\left( \frac{r}{r_{\rm eff}^{\rm intra}} \right) - Y_0\left( \frac{r}{r_{\rm eff}^{\rm intra}} \right) \right],
\end{equation}
\begin{equation}\label{eq:VinterLRreal}
	U_j^{\rm inter}(r) = \frac{\pi}{2\epsilon r_{\rm eff}^{\rm inter}}\left[H_0\left( \frac{r}{r_{\rm eff}^{\rm inter}} \right) - Y_0\left( \frac{r}{r_{\rm eff}^{\rm inter}} \right) \right].
\end{equation}
\end{subequations}
Here, $H_0$ and $Y_0$ are the zeroth Struve function and the Bessel function of the second kind, and we have defined $r_{\rm eff}^{\rm intra}=r_*^{(1)}+r_*^{(-1)}$ and $r_{\rm eff}^{\rm inter}=r_*^{(1)}+r_*^{(-1)}+d$. This corresponds to formula \eqref{eq:UK} in the main text.

\section{Matrix elements of the relative motion problem}\label{app:matrices}
Taking the basis states of \eqref{eq:basis}, the RM Hamiltonian matrix elements can be written as
\begin{equation}
	H_{j j'}^m = K_{j j'}^m - \frac{2\mu e^2}{\hbar^2\epsilon r_{\rm eff}}U_{j j'}^m,
\end{equation}
with auxiliary matrix elements
\begin{widetext}
\begin{subequations}
\begin{equation}
\begin{split}
	K_{j j'}^m \equiv& \int_0^\infty d\rho\,\rho\chi_j^{m *}(\rho) \left[\frac{\partial^2}{\partial \rho^2} + \frac{1}{\rho}\frac{\partial}{\partial \rho} - \frac{m^2}{\rho} \right] \chi_{j'}^m(\rho)=-[2|m|+1]\frac{\beta^{2|m|}\beta_j}{(\beta_j + \beta_{j'})^{2|m|+1}}\Gamma(2|m|+1)+\frac{\beta^{2|m|}\beta_{j}^2}{(\beta_j+\beta_{j'})^{2|m|+2}}\Gamma(2|m|+2),
\end{split}
\end{equation}
\begin{equation}
\begin{split}
	U_{j j'}^m \equiv& \frac{\pi}{2}\int_0^\infty d\rho\,\rho\chi_j^{m *}(\rho) \left[ H_0\left(\frac{\rho}{r_{\rm eff}}\right) - Y_0\left(\frac{\rho}{r_{\rm eff}}\right) \right]  \chi_{j'}^m(\rho)=\frac{\beta^{2|m|}}{r_{\rm eff}}\Bigg[ \frac{\Gamma(2|m|+3)}{\beta_{jj'}^{2|m|+3}} {}_3F_{2}\left(|m|+1,|m|+\tfrac{3}{2},|m|+2;\tfrac{3}{2},\tfrac{3}{2}; -r_{\rm eff}^{-2}\beta_{j j'}^{-2} \right)\\
	 &- 4^{|m|}r_{\rm eff}^{2|m|+3}\cos{(|m|\pi)}\Gamma^2(|m|+1) {}_2F_1\left(|m|+1,|m|+1;\tfrac{1}{2}; -r_{\rm eff}^2\beta_{jj'}^{2} \right) \Bigg],
\end{split}
\end{equation}
\end{subequations}
\end{widetext}
where ${}_pF_{q}(a_1,\ldots, a_p;b_1,\ldots,b_q;x)$ are generalized hypergeometric functions, and we have defined $\beta_{jj'}=\beta_j+\beta_{j'}$. Similarly, the RM overlap matrix elements are given by
\begin{equation}
\begin{split}
	S^m_{jj'}\equiv& \int_0^\infty d\rho\,\rho \chi_j^{m *}(\rho)\chi_{j'}^{m}(\rho) =\beta^{2|m|}\int_0^\infty d\rho\, \rho^{2|m|+1}\exp{-\beta_{jj'}\rho}\\
	=&\beta^{2|m|}\frac{\Gamma(2|m|+2)}{\beta_{jj'}^{2|m|+2}}.
\end{split}
\end{equation}

\section{Matrix elements of the center-of-mass problem}\label{app:melems}
The COM part of the Hamiltonian \eqref{eq:Heh} with the trigonally-warped harmonic potential \eqref{eq:fit} can be written in cylindrical coordinates as
\begin{equation}
	H_{\rm COM} = -\frac{\hbar^2}{2M}\left(\frac{\partial^2}{\partial R^2} + \frac{1}{R}\frac{\partial}{\partial R} + \frac{1}{R^2}\frac{\partial^2}{\partial \Phi^2}\right)+\frac{M\omega^2}{2}R^2 + \delta V,
\end{equation}
with the trigonal warping term
\begin{equation}
	V = \frac{M\omega^2}{2}R^2\cos{\left( 3\Phi + \varphi \right)}.
\end{equation}
Using the basis states \eqref{eq:basisCOM} (with the $x$ axis defined so as to make $\varphi=0$), which are eigenstates of the isotropic part of $H_{\rm COM}$, we get the matrix elements
\begin{equation}\label{eq:melemF}
\begin{split}
	&\braoket{\psi_{j',\mm'}}{H_{\rm COM}}{\psi_{j,\mm}} = \hbar \omega (j+1) \delta_{j',j}\delta_{\mm',\mm} \\
	&-\delta\frac{\hbar\omega}{4}\frac{V_{j',j}^{\mm',\mm}\delta_{|\mm-\mm'|,3}}{\sqrt{\left(\tfrac{j'-|\mm'|}{2}+1 \right)_{|\mm'|} \left(\tfrac{j-|\mm|}{2}+1 \right)_{|\mm|} }},
\end{split}
\end{equation}
where
\begin{equation}
\begin{split}
	V_{j',j}^{\mm',\mm}=&\frac{2}{R_0^4}\int_0^\infty dR\,R^3 \left(\frac{R}{R_0} \right)^{|\mm'|+|\mm|} \exp{-R^2/R_0^2}\\
	&\times L_{(j'-|\mm'|)/2}^{|\mm'|}(R^2/R_0^2)L_{(j-|\mm|)/2}^{|\mm|}(R^2/R_0^2).
\end{split}
\end{equation}
As discussed in the main text, the trigonal distortion of the potential leads to coupling between states with quantum numbers $\mm'=\mm\pm3$, conserving angular momentum only modulo 3. To evaluate the radial part, we change variables to obtain\cite{andrewsbook}
\begin{widetext}
\begin{equation}
\begin{split}
	&V_{j',j}^{\mm',\mm}=\int_0^\infty dx\,x^{\frac{|\mm'|+|\mm|}{2}+1} \exp{-x} L_{\tfrac{j'-|\mm'|}{2}}^{|\mm'|}(x)L_{\tfrac{j-|\mm|}{2}}^{|\mm|}(x)\\
	&=\frac{\Gamma\left(\tfrac{|\mm'|+|\mm|}{2}+\tfrac{5}{2} \right)\left(|\mm'|+1 \right)_{\tfrac{j'-|\mm'|}{2}} \left(|\mm|+1 \right)_{\tfrac{j-|\mm|}{2}}}{\left(\tfrac{j'-|\mm'|}{2} \right)!\left(\tfrac{j-|\mm|}{2} \right)!}
	\sum_{m=0}^{\tfrac{j'-|\mm'|}{2}}\frac{\left(\tfrac{|\mm'|-j'}{2}\right)_m\left(\tfrac{|\mm'|+|\mm|}{2}+\tfrac{5}{2} \right)_m}{(|\mm'|+1 )_m m!}\sum_{k=0}^{\tfrac{j-|\mm|}{2}}\frac{\left( \tfrac{|\mm|-j}{2}\right)_k \left(\tfrac{|\mm'|+|\mm|}{2}+\tfrac{5}{2}+m \right)_k}{(|\mm|+1)_k k!}.
\end{split}
\end{equation}
\end{widetext}

\section{Symmetry of the exciton wave function under $C_3$ rotations}\label{app:symmetry}
Let $u_{\alpha,\tau,s,\kk}(\rr)$ be the Bloch wave function of a $\tau\KK$-valley electron of band $\alpha$ and spin $s$, belonging to the symmetry group\cite{kormanyos} $C_{3h}$. This function transforms under $C_3$ rotations as $C_3 u_{\alpha,\tau,s,\kk}(\rr) = \phi_{\alpha,\tau}u_{\alpha,\tau,s,C_3\kk}(\rr)$, where the eigenvalue $\phi_{\alpha,\tau}$ depends on the symmetry point in the lattice about which the rotation is performed [metal atom, chalcogen atom or hollow site; see Eq.\ \eqref{eq:C3Bloch}]. Accordingly, the creation operator for such an electron transforms as
\begin{equation}\label{eq:C3cdagger}
	C_3 c_{\alpha,\tau,s}^\dagger(\kk) C_3^{-1} = \phi_{\alpha,\tau}c_{\alpha,\tau,s}^\dagger (C_3\kk).
\end{equation}
From the field operator definition
\begin{equation}
	\varphi_{\alpha,\tau,s}(\rr) = \sum_{\kk}\frac{e^{i(\tau\KK+\kk)\cdot\rr}}{\sqrt{S}}c_{\alpha,\tau,s}(\kk),
\end{equation}
we find that
\begin{equation}\label{eq:C3field}
	e^{-i\tau\KK\cdot\rr}C_3 \varphi_{\alpha,\tau,s}(\rr) C_3^{-1} = \phi_{\alpha,\tau}^*e^{-i\tau\KK\cdot C_3\rr}\varphi_{\alpha,\tau,s} (C_3\rr).
\end{equation}
Substituting \eqref{eq:C3field} into Eq.\ \eqref{eq:IXrealspace} gives
\begin{equation}
\begin{split}
	C_3|& {\rm IX}_{\tau',\tau}^{c',v;s} \rangle_{\bar{\mm},\ell}^{m,n}=\phi_{c',\tau'}\phi_{v,\tau}^*\int d^2r_e \int d^2r_h\,e^{i\tau' \KK'\cdot C_3\rr_{\rm e} - i\tau\KK\cdot C_3\rr_{\rm h}} \\
	&\times \Psi_{\bar{\mm},\ell}^{m,n}(\rr_{\rm e},\rr_{\rm h})\varphi_{c',\tau',s}^\dagger (C_3\rr_{\rm e}) \varphi_{v,\tau,s}(C_3\rr_{\rm h}) C_3| \Omega \rangle\\
	=&\phi_{c',\tau'}\phi_{v,\tau}^*\int d^2\bar{r}_e \int d^2\bar{r}_h\,e^{i\tau' \KK'\cdot \bar{\rr}_{\rm e} - i\tau\KK\cdot \bar{\rr}_{\rm h}} \\
	&\times \Psi_{\bar{\mm},\ell}^{m,n}(C_3^{-1}\bar{\rr}_{\rm e},C_3^{-1}\bar{\rr}_{\rm h})\varphi_{c',\tau',s}^\dagger (\bar{\rr}_{\rm e}) \varphi_{v,\tau,s}(\bar{\rr}_{\rm h}) C_3| \Omega \rangle,
\end{split}
\end{equation}
where $\bar{\rr}_{\rm e/h} = C_3\rr_{\rm e/h}$. From Eq.\ \eqref{eq:C3wf} we get $\Psi_{\bar{\mm},\ell}^{m,n}(C_3^{-1}\bar{\rr}_{\rm e},C_3^{-1}\bar{\rr}_{\rm h}) = C_3\Psi_{\bar{\mm},\ell}^{m,n}(\bar{\rr}_{\rm e},\bar{\rr}_{\rm h})=e^{-i\tfrac{2\pi}{3}(\bar{\mm}+m)}\Psi_{\bar{\mm},\ell}^{m,n}$. The many-body state $| \Omega \rangle$ must also be an eigenstate of the $C_3$ operator with some eigenvalue $\phi_{\Omega}$, resulting in Eq.\ \eqref{eq:C3IX}.

\section{The light-matter interaction Hamiltonian}\label{app:lightmatter}
In the second quantization formalism, the interaction between photons and Bloch electrons in a solid can be written as\cite{kira2011semiconductor}
\begin{equation}
\begin{split}
	H_{\rm LM} =& \frac{e}{2m_0c}\sum_{s,\eta}\sum_{\alpha,\beta}\sum_{\qq,\xxi}\sqrt{\frac{4\pi\hbar c}{SLq}}\langle \beta, \qq+\xxi | p_{-\eta} |\alpha,\qq \rangle\\
	 &c_{\beta,s}^\dagger(\qq+\xxi)c_{\alpha,s}(\qq)\left[ a_{\eta}(\xxi) + a_{-\eta}^\dagger(-\xxi) \right] + \text{H.c.},
\end{split}
\end{equation}
where $e$ and $m_0$ are the electron charge and mass, $p_{-\eta}=p_x -i\eta p_y$ is the helical momentum operator, and the sums over $\alpha$ and $\beta$ run over all bands of both layers. However, in the present case we shall be interested only in optical transitions between the conduction and valence bands within each layer, and between the overall highest valence band $v$ (of layer MX${}_2$) and the lowest conduction band $c'$ (of layer M'X'${}_2$). Moreover, we drop terms which simultaneously excite (relax) the electron and photon fields, since such processes do not conserve energy and cannot contribute to photon absorption or emission.

For intralayer optical transitions close to the $K$ valley in MX${}_2$ we have the matrix element $\langle c, \tau\KK+\kk+\xxi | p_{-\eta} |v,\tau\KK+\kk \rangle$, which may be approximated by its value exactly at the valley ($\kk=\xxi=0$), where the Bloch functions are symmetric under $C_3$ rotations\cite{liu2015electronic}:
\begin{equation}
\begin{split}
	\langle c, \tau\KK | p_{-\eta} |v,\tau\KK \rangle =& \langle c, \tau\KK |C_3^{-1}C_3 p_{-\eta} C_3^{-1}C_3|v,\tau\KK \rangle\\
	=& e^{i\tfrac{2\pi}{3}(\tau + \eta)} \langle c, \tau\KK | p_{-\eta} |v,\tau\KK \rangle.
\end{split}
\end{equation}
The matrix element is non zero only for $e^{i\tfrac{2\pi}{3}(\tau + \eta)}=1$, requiring $\eta = -\tau$, and the analogous result $\eta = -\tau'$ can be found for the M'X'${}_2$ layer. This is the well known optical selection rule for TMD monolayers. As a result, the intralayer light-matter interactions are given by
\begin{equation}
\begin{split}
	H_{\rm LM}^{\rm intra} =& \frac{e \gamma}{\hbar c}\sum_{s,\tau}\sum_{\kk,\xxi}\sqrt{\frac{4\pi\hbar c}{SLq}}c_{c,\tau,s}^\dagger(\kk+\xxi_\parallel)c_{v,\tau,s}(\kk)a_{-\tau}(\xxi) + \text{H.c.}\\
	&+\frac{e \gamma'}{\hbar c}\sum_{s,\eta}\sum_{\kk,\xxi}\sqrt{\frac{4\pi\hbar c}{SLq}}c_{c',\tau',s}^\dagger(\kk+\xxi_\parallel)c_{v',\tau',s}(\kk)a_{-\tau'}(\xxi) + \text{H.c.},
\end{split}
\end{equation}
where we have abbreviated $\gamma^{(')} = (\hbar/2m_0)\langle c^{(')}, \tau\KK | p_{\tau} |v^{(')},\tau\KK \rangle$, and specialized to the case of a 2D material by conserving only the in-plane component of the photon momentum.

Identifying allowed interlayer optical transitions is more subtle in the presence of a moir\'e pattern, since the structure has $C_3$ symmetry only about specific points $\{ \site\}$ in the moir\'e supercell. Labeling $C_3$ rotations about site $\site$ as $C_{3,\site}$ we get
\begin{equation}
\begin{split}
	\langle c', \tau'\KK' | p_{-\eta} |v,\tau\KK \rangle =& \langle c, \tau\KK |C_{3}^{-1}C_3 p_{-\eta} C_3^{-1}C_3|v,\tau\KK \rangle\\
	=& \phi_{c',\tau',\site}^*\phi_{v,\tau,\site}e^{i\tfrac{2\pi}{3}\eta} \langle c', \tau'\KK' | p_{-\eta} |v,\tau\KK \rangle,
\end{split}
\end{equation}
with $\site$-dependent eigenvalues $\phi_{c',\tau',\site}$ and $\phi_{v,\tau,\site}$ for the Bloch functions. Since we are interested in interlayer excitons localized at $\site=BA$ sites for $R$-type structures and $\site=BB'$ for $H$-type structures, we make use of the eigenvalues reported in Eq.\ \eqref{eq:C3Bloch} and obtain the conditions
\begin{subequations}
\begin{equation}
	e^{i\tfrac{2\pi}{3}(\eta - \tau)} = 1,\,(R\,\text{stacking}),
\end{equation}
\begin{equation}
	e^{i\tfrac{2\pi}{3}\eta} = 1,\,(H\,\text{stacking}).
\end{equation}
\end{subequations}
In the former case the resulting selection rule is $\eta = \tau$, whereas in the latter the equation cannot be fulfilled by in-plane circularly polarized photons ($\eta=\pm1$), forbidding direct interlayer optical transitions by valley carriers at BB' sites in $H$-stacked heterobilayers. These results were first reported in Ref.\ \onlinecite{hongyi2017}. It is worthwhile mentioning that localized valley exciton states do contain contributions from electrons and holes away from the valley, making the optical matrix element non zero but negligible.

To get an expression for the optical matrix element in the case of $R$-stacked structures we use the envelope function approximation for the Bloch functions, and Fourier expand them as
\begin{subequations}
\begin{equation}
	u_{v,\tau,s,\kk}(\rr,z) = \sum_{\GG}\frac{e^{i(\tau\KK+\GG+\kk)\cdot\rr}}{\sqrt{S}}u_{v,s}(\tau\KK+\GG+\kk,z),
\end{equation}
\begin{equation}
	u_{c',\tau,s,\kk'}(\rr,z) = \sum_{\GG'}\frac{e^{i(\tau\KK'+\GG'+\kk')\cdot(\rr-\rr_0)}}{\sqrt{S}}u_{c',s}(\tau\KK'+\GG'+\kk',z),
\end{equation}
\end{subequations}
where $\GG$ and $\GG'$ are Bragg vectors of the corresponding layers. Neglecting the photon momentum gives
\begin{equation}
\begin{split}
	&\langle c', \tau\KK'+\kk' | p_{-\tau} |v,\tau\KK+\kk \rangle \approx \sum_{\GG,\GG'}\delta_{\tau\KK'+\GG'+\kk',\tau\KK+\GG+\kk} e^{i\rr_0\cdot(\tau\KK+\GG)}\\
	&\times (-i\hbar)\int dz\, u_{c',s}^*(\tau\KK'+\GG',z)[i(\tau\KK+\GG)_{-\tau} + \partial_z]u_{v,s}(\tau\KK+\GG,z).
\end{split}
\end{equation}
Considering only Fourier components from the shortest possible Bragg vectors gives
\begin{equation}
\begin{split}
	&\langle c', \tau\KK'+\kk' | p_{-\tau} |v,\tau\KK+\kk \rangle \approx \sum_{\chi=0}^2\delta_{\kk-\kk',C_3^\chi \Delta \KK_{\tau,\tau}} e^{i\rr_0\cdot\tau C_3^\chi \KK}\\
	&\times (-i\hbar)\int dz\, u_{c',s}^*(\tau \KK',z)[i(\tau \KK)_{-\tau} + \partial_z]u_{v,s}(\tau \KK,z)\\
	&\equiv \gamma_R \sum_{\chi=0}^2\delta_{\kk-\kk',C_3^\chi \Delta \KK_{\tau,\tau}} e^{i\rr_0\cdot\tau C_3^\chi \KK}.
\end{split}
\end{equation}
Finally, the interlayer contribution for interaction between photons and charge carriers localized near the BA sites of $R$-stacked heterobilayers is
\begin{equation}
\begin{split}
	H_{\rm LM}^{\rm inter} =& \frac{e \gamma_R}{\hbar c}\sum_{s,\eta}\sum_{\kk',\kk,\xxi}\sqrt{\frac{4\pi\hbar c}{SLq}}\vartheta_{\tau,\tau}(\kk',\kk)\\
	&\times  c_{c',\tau,s}^\dagger(\kk'+\xxi_\parallel)c_{v,\tau,s}(\kk)a_{\tau}(\xxi) + \text{H.c.},
\end{split}
\end{equation}
where $\vartheta_{\tau',\tau}(\kk',\kk)$ has been defined in Eq.\ \eqref{eq:vartheta}. The full light-matter Hamiltonian is then given as $H_{\rm LM} = H_{\rm LM}^{\rm intra}+H_{\rm LM}^{\rm inter}$.

Note that each term in $H_{\rm LM}$ commutes with the $C_{3,\site}$ operators. We can see this for $R$-type structures ($\site=BA$) by using
\begin{subequations}
\begin{equation}
	C_{3,BA} c_{c',\tau,s}^\dagger(\kk) C_{3,BA}^{-1} = \phi_{c',\tau,BA} c_{c',\tau,s} = e^{-i\tfrac{2\pi}{3}\tau} c_{c',\tau,s}(C_3\kk),
\end{equation}
\begin{equation}
	C_{3,BA} c_{v,\tau,s}(\kk) C_{3,BA}^{-1} = \phi_{v,\tau,BA}^* c_{v,\tau,s} = e^{-i\tfrac{2\pi}{3}\tau} c_{v,\tau,s}(C_3\kk),
\end{equation}
\begin{equation}
	C_{3,BA} a_{\tau}(\xxi) C_{3,BA}^{-1} = e^{-i\tfrac{2\pi}{3}\tau} a_{\tau}(C_3 \xxi),
\end{equation}
\end{subequations}
to evaluate
\begin{equation}
\begin{split}
	&C_{3,BA} H_{\rm LM}^{\rm inter} C_{3,BA}^{-1} = \frac{e \gamma_R}{\hbar c}\sum_{s,\eta}\sum_{\kk',\kk,\xxi}\sqrt{\frac{4\pi\hbar c}{SLq}}\vartheta_{\tau,\tau}(\kk',\kk)\\
	&\times  c_{c',\tau,s}^\dagger(C_3\kk'+C_3\xxi_\parallel)c_{v,\tau,s}(C_3\kk)a_{\tau}(C_3\xxi)\left(e^{-i\tfrac{2\pi}{3}\tau} \right)^3 + \text{H.c.}\\
	&= H_{\rm LM}^{\rm inter}.
\end{split}
\end{equation}
The same holds for $H_{\rm LM}^{\rm intra}$.

\section{Further moir\'e localized states}\label{app:morelevels}
\begin{figure*}[h!]
\begin{center}
\includegraphics[width=1.6\columnwidth]{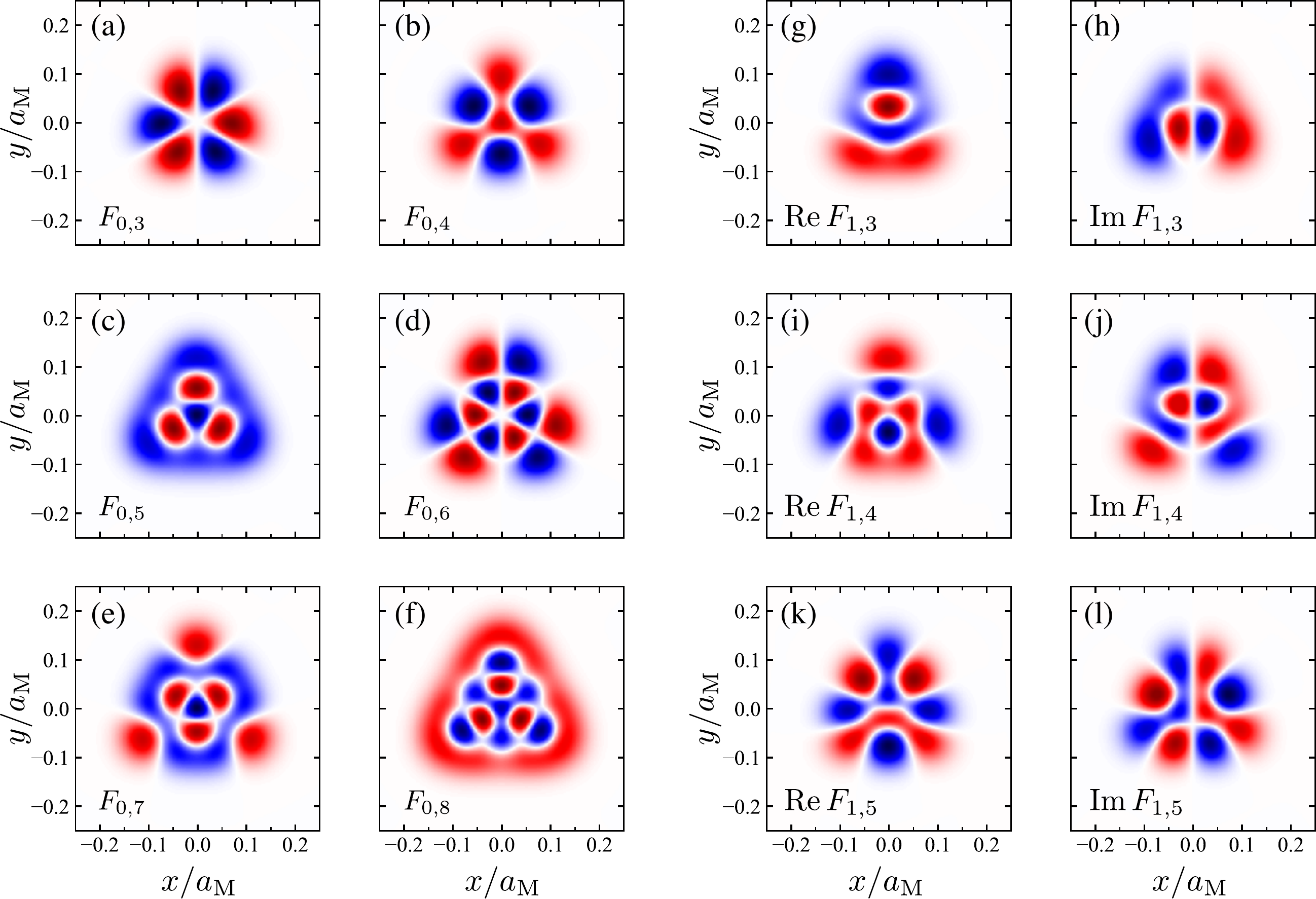}
\caption{Wavefunctions of moir\'e bound states in $R$-WSe${}_2$/MoSe${}_2$. As in Fig.\ \ref{fig:RWSe2MoSe2wavefunction}, each function has been normalized to its maximum value for illustration purposes, and the color blue (red) indicates positive (negative) values, whereas white means a zero value. (a)-(f) $F_{0,\ell}$ for $3\le \ell \le 8$. For $\ell=4,\,5,\,7$ and $8$ the functions are even under the mirror operations $\sigma_v$, $\sigma_v'$ and $\sigma_v''$ shown in Fig.\ \ref{fig:RWSe2MoSe2wavefunction}(c), and belong to irrep $A_1$ of point group $C_{3v}$. By contrast, $F_{0,3}$ and $F_{0,6}$ belong to irrep $A_2$, and are odd under the same mirror operations. (g)-(l) $F_{1,\ell}$ for $\ell \le 3 \le 5$, all of which belong to the two-dimensional irrep $E$. The corresponding wave functions from block $\bar{\mm}=-1$ can be obtained by complex conjugation.}
\label{fig:HighEnergyStates}
\end{center}
\end{figure*}

Figure \ref{fig:HighEnergyStates} shows the COM wave functions of states $F_{\bar{\mm},\ell}$ for $(\bar{\mm}=0,\,\ell=3,4,5,6,7,8)$ and $(\bar{\mm}=\pm1,\,\ell=3,4,5)$. States $F_{0,4},\,F_{0,5},\,F_{0,7}$ and $F_{0,8}$ belong to irrep $A_1$ of group $C_{3v}$, whereas $F_{0,3},\,F_{0,6}$ belong to irrep $A_2$. This can be inferred from the former's invariance under $\sigma_v$ mirror operations, the latter's acquisition of a $(-1)$ phase factor under the same, and all states' invariance under $C_3$ rotations. States $F_{\pm1,3}$, $F_{\pm1,4}$ and $F_{\pm1,5}$ belong to the two dimensional irrep $E$. We have numerically verified their properties under $C_3$ rotations and $\sigma_v$ mirror transformations.

\bibliographystyle{apsrev4-1}
\bibliography{references}
\end{document}